\documentclass[lettersize,journal]{IEEEtran}
\usepackage{amsmath,amsfonts}
\usepackage{autobreak}
\usepackage{algorithm}
\usepackage{algorithmicx}
\usepackage{algpseudocode}

\usepackage{array}
\usepackage{textcomp}
\usepackage{stfloats}
\usepackage{url}
\usepackage{verbatim}
\usepackage{graphicx}
\usepackage{cite}
\usepackage{subfigure}
\usepackage{float}
\usepackage{color}
\usepackage[T1]{fontenc}

\begin{document}
		
		\title{Networked Integrated Sensing and Communications for 6G Wireless Systems}
		
		\author{{Jiapeng Li, Xiaodan Shao,~\IEEEmembership{Member,~IEEE}, Feng Chen, \IEEEmembership{Member, IEEE}, Shaohua Wan, \IEEEmembership{Senior Member, IEEE}, Chang Liu, \IEEEmembership{Member, IEEE}, Zhiqiang Wei, \IEEEmembership{Member, IEEE}, and  Derrick Wing Kwan Ng, \IEEEmembership{Fellow, IEEE}}
			
			\thanks{	
			The work of C. Liu was supported by La Trobe University Research Project under Grant 325011150. 
			The work of Z. Wei was supported in part by the National Key R\&D Program of China (No. 2023YFA1008600). (Corresoonding authors: Xiaodan Shao; Feng Chen)
			
			J. Li, and F. Chen are with the College of Artificial Intelligence, Southwest University, Chongqing, 400715, China, the Brain-Inspired Computing and Intelligent Control Key Laboratory, Southwest University, Chongqing 400715, China, and the Key Laboratory of Luminescence Analysis and Molecular Sensing, Southwest University, Chongqing, 400715, China (e-mail: {\tt ljpskn@email.swu.edu.cn, fengchenmit@swu.edu.cn}).
			
			X. Shao is with the College of Artificial Intelligence, Southwest University, Chongqing 400715, China (e-mail: {\tt shaoxiaodan@zju.edu.cn}).
				
			S. Wan is with the Shenzhen Institute for Advanced Study, University of Electronic Science and Technology of China, Shenzhen 518110, China  (e-mail: {\tt shaohua.wan@ieee.org}).
			
			Z. Wei is with the School of Mathematics and Statistics, Xi'an Jiaotong University, Xi'an 710049, China, and also with the Peng Cheng Laboratory, Shenzhen, Guangdong 518055, China, and also with the Pazhou Laboratory (Huangpu), Guangzhou, Guangdong 510555, China (e-mail: {\tt zhiqiang.wei@xjtu.edu.cn}).
			
			C. Liu is with the Department of Computer Science and Information Technology, La Trobe University, Melbourne, Victoria 3150, Australia (e-mail: {\tt C.Liu6@latrobe.edu.au}).
			
			D. W. K. Ng is with the School of Electrical Engineering and Telecommunications, University of New South Wales, Sydney, NSW 2052, Australia (e-mail: {\tt w.k.ng@unsw.edu.au}). 
		}}

		\maketitle
		
		\begin{abstract}
			
			Integrated sensing and communication (ISAC) is envisioned as a key pillar for enabling the upcoming sixth generation (6G) communication systems, requiring not only reliable communication functionalities but also highly accurate environmental sensing capabilities.
			In this paper, we design a novel networked ISAC framework to explore the  collaboration among multiple users for environmental sensing.
			Specifically, multiple users can serve as powerful sensors, capturing back scattered signals from a target at various angles to facilitate reliable computational imaging.
			Centralized sensing approaches are extremely sensitive to the capability of the leader node because it requires the leader node to process the signals sent by all the users.
			To this end, we propose a two-step distributed cooperative sensing algorithm that allows low-dimensional intermediate estimate exchange among neighboring users, thus eliminating the reliance on the centralized leader node and improving the robustness of sensing.
			This way, multiple users can cooperatively sense a target by exploiting the block-wise environment sparsity and the interference cancellation technique.
			Furthermore, we analyze the mean square error of the proposed distributed algorithm as a networked sensing performance metric and propose a beamforming design for the proposed network ISAC scheme to maximize the networked sensing accuracy and communication performance subject to a transmit power constraint.
			Simulation results validate the effectiveness of the proposed algorithm compared with the state-of-the-art algorithms.
		\end{abstract}
		
		\begin{IEEEkeywords} Networked sensing, integrated sensing and communication, computational imaging, 6G, beamforming design.
		\end{IEEEkeywords}

		\section{Introduction}

		The sixth-generation (6G) wireless networks are expected to enable numerous emerging applications such as extending reality\cite{Extended_reality_Alizad}, holographic communications\cite{Holographic_com_AI_Strinati}, smart-health medical\cite{Medical_AI_Shen}, smart cities\cite{Internet_of_things_Zanella}, and autonomous driving\cite{Intelligent_Toutouh}. These applications require an enhanced wireless communication capability as well as accurate environmental information such as the location, shape, and electromagnetic characteristics of the targets and/or background scatters in the environment\cite{Shaonwad150, shao2022target,shao2022magazine}. Meanwhile, the extensive use of  high-frequency communication technologies has enabled high-resolution wireless signals in both space and time, offering the potential for high-performance sensing\cite{8789678,8408563,8798636}. To address this need, a new technology paradigm called integrated sensing and communications (ISAC) has been developed. ISAC seamlessly integrates two originally decoupled functionalities into one single unique system by exploiting existing wireless communication devices and infrastructures to achieve effective environment sensing \cite{cui2021integrating,wei_isac,li2023beam}. In this manner, wireless sensing improves the utilization of system resources efficiency by sharing the same frequency, hardware, protocol, and network with wireless communications synergically \cite{mao, ISAC_in6G, ISAC_chang}. Both practical radio sensing and communication systems have been developed toward adopting higher frequency bands and larger antenna arrays which provides a fascinating opportunity to exploit the upcoming 6G wireless infrastructure for the realization of ISAC.

		In recent years, ISAC has attracted significant research attention in various application scenarios. For instance, in \cite{wang2022noma}, an ISAC framework for non-orthogonal multiple access was designed to maximize the effective sensing power and the communication throughput.
		Besides, to guarantee low communications latency and accurate beam tracking, the authors in \cite{mu2021integrated} proposed an ISAC-enabled predictive beamforming scheme.
		Also, in \cite{zhang2022accelerating}, the authors investigated the
		acceleration of edge intelligence via ISAC.
		Furthermore, to improve the	per-user signal-to-interference-plus-noise ratio (SINR) and/or the total target illumination power, the integration of an intelligent reflecting surface into an ISAC system was considered recently in several works, e.g.,\cite{10041837,9733335, 10229204}.
		Thanks to its flexibility, ISAC is expected to have wide applications in smart homes, smart manufacturing, environmental monitoring, extended reality
		and so on. This calls for the deployment of a large number of wireless devices such as access points (AP)/base stations (BS) and mobile terminals that naturally serve as an excellent foundation \cite{wild2021joint}, paving the way for the development of networked ISAC. However, the collaboration among wireless devices for cooperative sensing has not been explored yet due to the lack of a unified networked sensing protocol. Indeed, effectively exploiting the inherent network characteristics for wireless sensing and resource allocation remains an open problem.

		One major issue with sensing is the occlusion effect among objects and blockages in propagating electromagnetic waves, which hinders the reception of echoes from sensing targets. The situation is particularly challenging when only a single user is adopted to sense the entire object. In this paper, we propose a novel networked ISAC framework in which multiple users in the network can participate together in different locations for joint sensing. Different from the work mentioned above, our goal is to design an ISAC system based on existing wireless systems that exploits the collaboration among multiple users in different locations for environmental sensing (imaging). Meanwhile, the potentially enormous number of unknown variables caused by the environment renders the sensing task under-determined and eventually intractable. Therefore, the inherent environment sparsity can be exploited to address this issue. In this paper, we consider the intrinsic sparsity of objects within an environment in the channel for sensing. To the best of our knowledge, related existing literature is still in its infancy.

		In this paper, we consider a downlink far-field scenario with multiple users. Specifically, a target scatters the downlink signals from a BS to multiple users, which can be exploited and adopted to jointly reconstruct the wireless image of the target by estimating the radio wave propagation characteristics \cite{JRAC_Liu}, \cite{imaging_3}. In this scenario, the proposed networked-ISAC serves as a handy tool to facilitate the exploration of the potential gain in cooperative sensing. In this paper, we first design a two-step scheme to reduce interference from communication symbols to the sensing module. Then, we formulate target sensing as an imaging problem and propose a centralized sensing resolution for the multi-user-assisted ISAC system where global data is available for one leader user to unveil important insights for system design. As for the practical case where only local data is available at each user, we design a distributed cooperative sensing strategy. Without the need for having a leader user and fusion center, involved users in cooperative sensing are required only to exchange intermediate estimates with neighboring nodes. Thereby all users in the system can obtain final sensing results, improving system robustness compared with centralized counterpart. Additionally, our distributed sensing strategy takes into account both input noise and environment noise to improve sensing performance.

		The contributions of this paper are summarized as follows:
		\begin{itemize}
			
			\item We propose a unified framework for networked ISAC in 6G wireless networks to improve spectral efficiency and sensing accuracy.
			Based on the proposed framework, we design a two-step distributed sensing algorithm where interference caused by data symbols to sensing and block-sparsity of the environment are jointly taken into account to improve sensing accuracy.

			\item We analytically characterize the stability property of the proposed algorithm’s convergence and derive the networked sensing performance metric. Additionally, we reveal the impact of key parameters on the proposed distributed sensing scheme and overall performance of ISAC, providing valuable guidance for practical ISAC design.
			
			\item We design a beamforming optimization algorithm for the proposed networked ISAC scheme based on the derived performance metric. The algorithm is designed to minimize imaging error and maximize communication quality subject to a transmit power constraint. Numerical results demonstrate the superiority of our design.
			
		\end{itemize}
		
		The remainder of this paper is organized as follows.
		We first introduce the system model of the ISAC-assisted imaging system in Section \ref{section_2}.
		Then, we design the sensing algorithm and analyze the performance in Section \ref{section_3}.
		After that, we formulate the beamforming optimization problem for the proposed networked ISAC scheme and present the methodology for obtaining a numerical solution in Section \ref{section_4}.
		Numerical results are presented in Section \ref{section_5}.
		Finally, we conclude this paper in Section \ref{section_6}.
		
		\emph{Notations}: We adopt regular font letters for scalars, bold lowercase letters for vectors, and bold uppercase letters for matrices.
		We use $\operatorname{col}\{\cdot\}$ to denote a column vector, $\mathbf{I}_{K}$ to denote the $K \times K$ identity matrix, $\|\cdot\|_{p}$ to denote $\ell_p$-norm, $\operatorname{diag}\{\cdot\}$ to denote the (block) diagonal matrix, $\mathcal{C} \mathcal{N}\left(0, \sigma^{2} \right)$ to denote the complex Gaussian distribution with mean 0 and variance $\sigma^{2}$,
		and $|\cdot|$ to denote the absolute value of a scalar.
		$\mathbb{C}^{m \times n}$ represents the sets of an $m$-by-$n$ dimensional complex matrix.
		Operators  $E[\cdot]$, $\operatorname{vec}\{\cdot\}$, $(\cdot)^H$, $(\cdot)^T$, $(\cdot)^\dag$, $\operatorname{tr}(\cdot)$, and $\operatorname{Rank}(\cdot)$ denote the  expectation, vectorization, conjugate transpose, transpose, complex conjugate, the trace, and the rank of a matrix, respectively.
		$\otimes$ denotes Kronecker product of two matrices.
		
		\section{System Model}\label{section_2}
		\begin{figure*}[t]
			\setlength{\abovecaptionskip}{-0.cm}
			\setlength{\belowcaptionskip}{0.cm}
			\centering
			\includegraphics [width=0.73\textwidth] {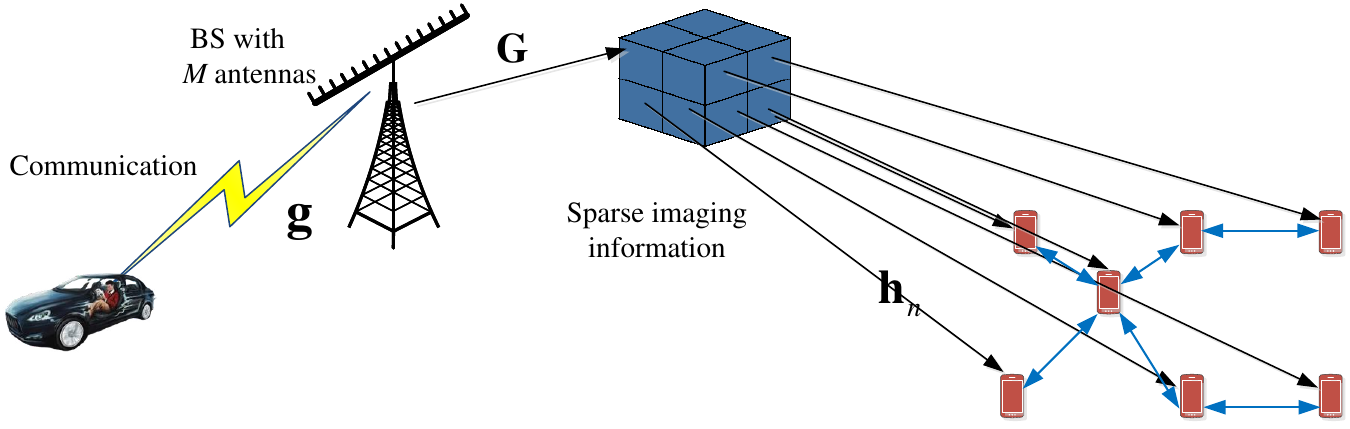}
			\caption{The proposed networked ISAC systems with a target, $N$ sensing users, and one communication user.}
			\label{FL}
		\end{figure*}
\begin{figure}[t]
			\setlength{\abovecaptionskip}{-0.2cm}
			\setlength{\belowcaptionskip}{0.2cm}
			\centering
			\includegraphics [width=0.45\textwidth,trim = 20 20 20 20,clip] {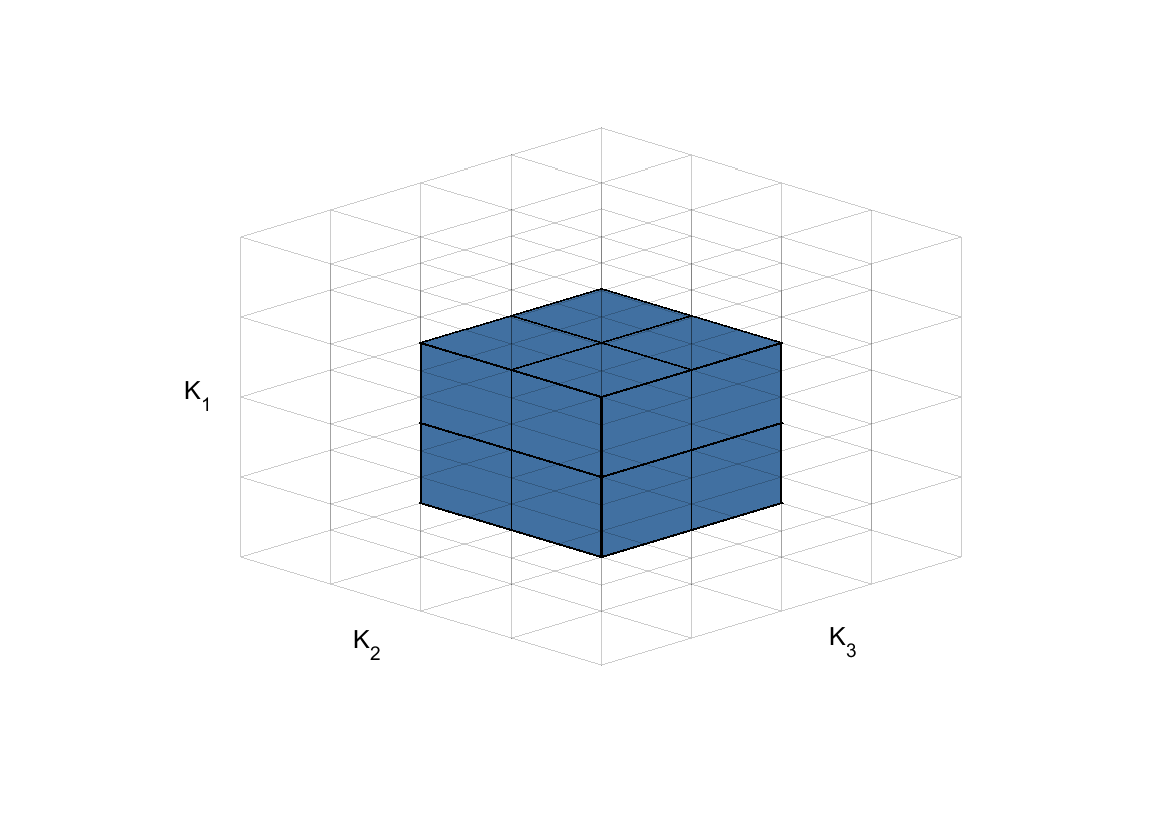}
			\caption{ Discretizing environmental targets into multiple sub-cubes of ROI for effective sensing.}
			\label{ROI}
		\end{figure}

		We consider a wireless network consisting of a BS equipped with $M$ antennas that senses multiple single-antenna users. As shown in Fig. \ref{FL}, in the downlink communication scenario, the electromagnetic wave sent by the BS is reflected by the target and then collected by $N$ sensing users. Meanwhile, the BS maintains its communication with another communication user. In this paper, we focus on multi-user collaborative sensing for computational imaging. We consider 3D imaging based on pixel division for ISAC that divides the sparse region of imaging (ROI) about a target into pixels \cite{JCAS_IRS_multi_user_Tong}, \cite{imaging_2}, as shown in Fig. \ref{ROI}. In other words, the ROI is divided into a number of sub-cubes of equal sizes where each sub-cube is regarded as a pixel and a target usually consists of multiple pixels.  In practical scenarios, the size of each pixel is determined by the size of the ROI and the wavelength of the signal.
		In this paper, we assume all sub-cubes in the ROI can be illuminated by the scattered signal, a common assumption in sensing literature \cite{shao,JCAS_IRS_multi_user_Tong}.
		Mathematically, the scattering coefficient of the ROI is characterized by a sparse vector ${\bf{x}} = {[{x_1}, {x_2},\cdots, {x_K}]^T}$, where ${x}_k \ge 0,  \forall k \in \{ 1, \dots, K \}$, represents the scattering amplitude of the $k$-th pixel. If the $k$-th pixel is empty, then ${x}_k = 0$.
		Assume that the number of pixels in the length, width, and height dimensions of the ROI is $K_1$, $K_2$, and $K_3$ respectively, then we have $K = K_1 \times K_2 \times K_3$.
		The number of pixels determines the accuracy limit of the sensing capability; the more the number, the more accurate the estimation is possible.
		Without loss of generality, we assume that channel state information is known to all users. Moreover, it is assumed that the distance between the communication user and the target is far away such that the signal from the BS to the target will not be reflected back to the communication user.
		We denote $\mathbf{g} \in \mathbb{C}^{M\times 1} $, $\mathbf{G} \in \mathbb{C}^{M \times K}$, $\mathbf{h}_{n} \in \mathbb{C}^{K\times 1}, \forall n \in \{1,\dots,N\} $ as the channel from the BS to the communication device, the channel from the BS to the ROI, and the channel from the ROI to the $n$-th sensing user, respectively.
		
		If we adopt a uniform sampling with a rate of $f_s$ at the transmitter and the receiver, the number of sampling points is $T=\tau \times f_s$ with $\tau \geq 0$ being the time slot duration.
		In practical situations, communication and sensing, as two essential functions of the BS, are often required by users at the same time.
		Similar to \cite{zhang2022accelerating}, we assume that both the information signals and the sensing signals are simultaneously transmitted for communication and sensing.
		The transmitted signals at the BS can be expressed as
		\begin{align}\label{tr}
			{\bf{s}}_{i}=\mathbf{w}{{s}}_{i}^{\mathrm{e}}+\mathbf{f}{{s}}_{i}^{\mathrm{d}},
		\end{align}
		where $\mathbf{w} \in \mathbb{C}^{M\times 1}$ and $\mathbf{f} \in \mathbb{C}^{M \times 1}$ denote the sensing beamformer and the data stream beamformer, respectively, $s_{i}^{\mathrm{e}} \in \mathbb{C}$ and $s_{i}^{\mathrm{d}} \in \mathbb{C}$ are the sensing and data symbols at time slot $i, i\in \{1,\dots,T\}$, respectively.
		For ease of analysis, it is assumed that the sensing signals and the communication signals are mutually independent and satisfy $E[s_i^{\mathrm{e}}\left(s^{\mathrm{d}}_i\right)^H]=0$.
		In the considered system, the multiple users in the environment share the same time-frequency resources, similar to \cite{li2022integrated}.
		Therefore, the received signal at the $n$-th sensing user at time $i$ is given by
		\begin{align}\label{se4}
			y_{n,i}= \left(s_{i}^{\mathrm{e}}\mathbf{w}^H{\bf{G}}{\rm{diag}}({{\bf{h}}_n})+s_{i}^{\mathrm{d}}\mathbf{f}^H{\bf{G}}{\rm{diag}}({{\bf{h}}_n})\right){\bf{x}}+{v_{n,i}},
		\end{align}
		where $v_{n,i}$ denotes the additive white Gaussian noise (AWGN) at the $n$-th user and follows $\mathcal{C} \mathcal{N}\left(0, \sigma_{\mathrm{o}}^{2} \right)$ with $\sigma_{\mathrm{o}^{2}}$ being the noise variance.
		By stacking all the signals received by the $n$-th user over the duration $T$ into a vector $\mathbf{y}_n$, we obtain
		\begin{align}\label{se3}
			\mathbf{y}_{n}=  \left(\mathbf{s}^{\mathrm{e}}\mathbf{w}^H{\bf{G}}{\rm{diag}}({{\bf{h}}_n})+\mathbf{s}^{\mathrm{d}}\mathbf{f}^{H}{\bf{G}}{\rm{diag}}({{\bf{h}}_n})\right){\bf{x}}+{\bf{v}}_{n},
		\end{align}
		where $\mathbf{y}_{n}=[{y}_{n,1},\cdots,{y}_{n,T}]^{T}$, $\mathbf{v}_{n}=[{v}_{n,1},\cdots,{v}_{n,T}]^{T}$,  $\mathbf{s}^{\mathrm{e}}=[{s}^{\mathrm{e}}_1,\cdots,{s}^{\mathrm{e}}_T]^{T}$, and $\mathbf{s}^{\mathrm{d}}=[{s}^{\mathrm{d}}_1,\cdots,{s}^{\mathrm{d}}_T]^{T}$.
		
		On the other hand, the signal received by the communication user at time $i$ is given by
		\begin{align}\label{ed}
			y^{i}_{\mathrm{es}}=\mathbf{g}^H\mathbf{w}{{s}}_{i}^{\mathrm{e}}  +\mathbf{g}^H\mathbf{f}{{s}}_{i}^{\mathrm{d}}+\bar{z}_{i},
		\end{align}
		where $\bar{z}_i$ denotes the AWGN and follows $\mathcal{C} \mathcal{N}\left(0, \sigma_{\mathrm{o}}^{2} \right)$.
		In this paper, we aim to design a sensing algorithm for reconstructing $\mathbf{x}$ and optimize the beamforming strategy for sensing and communication.
		
		\section{Sensing Algorithm Design} \label{section_3}
		In this section, we first design a two-step sensing scheme that exploits the interference cancellation technique to reduce the impact of unknown data symbols for sensing. Then, we design a centralized sensing strategy for wireless imaging in the considered ISAC system. All the sensing computations are processed at a leader node. Afterwards, we extend the centralized imaging strategy to a distributed approach while taking into account some practical limitations of the centralized strategy. Then, we analyze the theoretical performance metric of the algorithm for quantifying distributed sensing errors.

		\subsection{Two-Step Sensing Algorithm}
		For the purpose of sensing, the ISAC system calculates the ROI vector for sensing based on the received signal by the sensing users. Unlike known $s^{\mathrm{e}}$, the data symbol $s^{\mathrm{d}}$ is unknown to the sensing users and does not play an active role for imaging in sensing.
		If the data symbols are simply regarded as noise, the estimates of the ROI vector would experience a significant degradation in sensing performance \cite{ISAC_in6G}, \cite{9540344}.
		Therefore, we propose a two-step sensing algorithm to reduce the impacts caused by interference from $s^{\mathrm{d}}$. The main idea of our proposed two-step sensing algorithm is to compensate for the deviation in the second step based on the estimated parameters in the first step.

		We now provide the details of the proposed two-step algorithm. In the first step, we treat the unknown data symbols as noise and directly estimate the ROI vector by utilizing the sensing symbols to obtain a coarse estimation. In this case, we set the signal received by the $n$-th sensing user at time $i$ as
		\begin{align}\label{sens}
			{{y}}_{n,i} = \left(\underbrace{s_{i}^{\mathrm{e}}\mathbf{w}^H{\bf{G}}{\rm{diag}}({{\bf{h}}_n})+\mathbf{e}_{n,i}}_{\mathbf{\tilde{u}}^{(1) H}_{n,i}}\right)\mathbf{x}+{v}_{n,i},
		\end{align}
		where $\mathbf{e}_{n,i}={s}^{\mathrm{d}}_i\mathbf{f}^{H}{\bf{G}}{\rm{diag}}({{\bf{h}}_n})$, which is considered as the input noise of the sensing module.
		Therefore, we can obtain an estimate $\hat{\mathbf{x}}$ of the ROI vector in the first step by taking $({{y}}_{n,i},\mathbf{\tilde{u}}^{(1)}_{n,i})$ as the input for the sensing module, where $\mathbf{\tilde{u}}^{(1)}_{n,i} = \left(  s_{i}^{\mathrm{e}}\mathbf{w}^H{\bf{G}}{\rm{diag}}({{\bf{h}}_n})+\mathbf{e}_{n,i}\right)^{H} $.
		Algorithm \ref{alg:algorithm_SDTLS} in Section III-C will provide a detailed description of a specific estimation algorithm. 
		However, this estimate is biased due to the fact that the first step ignores the impact of the unknown data symbols.
		
		The component containing the data symbols is then calculated as the difference between the received signal and the  estimates of the $n$-th user, i.e.,
		\begin{align} \label{sen6}
			\hat{\mathbf{y}}^{(1)}_{n} = \mathbf{y}_{n} - \mathbf{s}^{\mathrm{e}}\mathbf{w}^H{\bf{G}}{\rm{diag}}({{\bf{h}}_n})\hat{\mathbf{x}} , \forall n,
		\end{align}
		which equals to the estimation of the unknown data term $\mathbf{s}^{\mathrm{d}}\mathbf{f}^{H}{\bf{G}}{\rm{diag}}({{\bf{h}}_n}){{\bf{x}}}+{\bf{v}}_{n}$ in the sensing model (\ref{se3}).
		
		Since the obtained $\hat{\bf{x}}$ in the first step is biased, it is expected that there is still a large error for directly estimating the data symbols $\mathbf{s}^{\mathrm{d}}$.
		In order to compensate for the error caused by the data symbol $\mathbf{s}^{\mathrm{d}}$, we construct a sparse vector $\mathbf{z}_{n} \in \mathbb{C}^{T\times 1} $ for each user. Then, by ignoring the effect of the noise terms ${\bf{v}}_{n}$ at the moment, we have
		\begin{equation}
		\begin{aligned}\label{ft}
			\hat{\mathbf{Y}}_{n}^{(1)}=  \hat{\mathbf{y}}_n^{(1)} \mathbf{z}_{n}^{H}&=(\mathbf{y}_{n} - \mathbf{s}^{\mathrm{e}}\mathbf{w}^H{\bf{G}}{\rm{diag}}({{\bf{h}}_n})\hat{\mathbf{x}})\mathbf{z}_n^{H} \\
			& \approx \mathbf{s}^{\mathrm{d}}\mathbf{f}^{H}{\bf{G}}{\rm{diag}}({{\bf{h}}_n}){{\bf{x}}}\mathbf{z}_n^{H}.
		\end{aligned}
		\end{equation}
		If we replace $\mathbf{f}^{H}{\bf{G}}{\rm{diag}}({{\bf{h}}_n})\hat{{\bf{x}}}$ with an auxiliary variable $\epsilon_n \in \mathbb{C}$, we can rewrite \eqref{ft} as $\hat{\mathbf{Y}}_{n}^{(1)} = \epsilon_n \mathbf{s}^{\mathrm{d}}\mathbf{z}_{n}^{H}$ and $\text{rank}(\hat{\mathbf{Y}}_{n}^{(1)}) = 1$.
		Therefore, by performing singular value decomposition (SVD) on ${\hat{\mathbf{Y}}}_{n}^{(1)}$ yields ${\hat{\mathbf{Y}}}_{n}^{(1)} = \mathbf{Q} \mathbf{\Sigma} \mathbf{P} $, where $\mathbf{\Sigma}\succeq \mathbf{0}$ is a diagonal matrix consisting of the singular values of $\hat{\mathbf{Y}}_{n}^{(1)}$ sorted from the largest to the smallest, the data symbols in the second step can be estimated as $\hat{\mathbf{s}}^{\mathrm{d}} = \mathbf{Q}(:,1)$, which is the first column vector of $\mathbf{Q}$.
		
		However, since \eqref{ft} ignores the effect of noise term ${\bf{v}}_{n}$, there is still some error between the estimated $\hat{\mathbf{s}}^{\mathrm{d}}$ and the actual ${\mathbf{s}}^{\mathrm{d}}$.
		Thus, after estimating the ROI and the data symbols, the residual received  signal can be obtained by
		\begin{equation}
		\begin{aligned}\label{see3}
			\tilde{y}_{n,i}&={y}_{n,i}-\hat{\mathbf{s}}^{\mathrm{d}}_i\mathbf{f}^{H}{\bf{G}}{\rm{diag}}({{\bf{h}}_n})\hat{\mathbf{x}}\\ &=  \left(\underbrace{{s}^{\mathrm{e}}_i\mathbf{w}^H{\bf{G}}{\rm{diag}}({{\bf{h}}_n})+\boldsymbol{\varpi}_n}_{\mathbf{\tilde{u}}^{(2) H}_{n,i}}\right){\bf{x}}+{v}_{n,i},
		\end{aligned}
		\end{equation}
		where $ \mathbf{\tilde{u}}^{(2)}_{n,i} = \left( {s}^{\mathrm{e}}_i\mathbf{w}^H{\bf{G}}{\rm{diag}}({{\bf{h}}_n})+\boldsymbol{\varpi}_n\right)^{H}$ is the input for sensing in the second step, and ${\boldsymbol{\varpi}}_n=[{\varpi}_{n,1},\cdots,{\varpi}_{n,K}]^T \in \mathbb{C}^K$ is the residual interference caused by the imperfect data symbol estimation, which can usually be considered as AWGN.
		After that, by taking $(\tilde{y}_{n,i},\mathbf{\tilde{u}}^{(2)}_{n,i})$ as the input for the sensing module, we can re-estimate $\hat{\mathbf{x}}$ in the second step.
		Owing to the interference cancellation performed in the second step, the power of the ${\boldsymbol{\varpi}}_{n,i}$ in \eqref{see3} is much lower than the power of $\mathbf{e}_{n,i}$ in \eqref{sens}.
		The proposed two-step algorithm can obtain an estimate of the ROI vector while reducing the impact of unknown data symbols on the sensing performance, as will be demonstrated by the numerical examples in Section \ref{section_5}.

		For clarity, the pseudo-code of the proposed algorithm is outlined in Algorithm \ref{alg:algorithm_SDTLS_two_step}. In the first step (line 1-2), biased parameter estimation is conducted using the input $({y}_{n,i},\mathbf{\tilde{u}}^{(1)}_{n,i})$ obtained from the received signal in (\ref{sens}).
		In the second step (line 3-7), the algorithm obtains the input $(\tilde{y}_{n,i},\mathbf{\tilde{u}}^{(2)}_{n,i})$ from \eqref{see3} and re-estimates the parameters. The two-step parameter estimation in Algorithm 1 produces an unbiased estimate of the ROI.

		\begin{algorithm}[t]
			\caption{The Proposed Two-Step Distributed Sensing Algorithm}  \label{alg:algorithm_SDTLS_two_step}
			
			\textbf{Inputs:}
			$\mathbf{y}_{n},   {\bf{G}}, \mathbf{s}^{\mathrm{e}},   \mathbf{s}^{\mathrm{d}}, {\bf{h}}_n, \mathbf{w}, \mathbf{f}$

			\begin{algorithmic}[1]
				\Statex {\!\!\!\!\!\!\!\!\!\!\textbf{Step 1:}}
				
				\State {$\mathbf{\tilde{u}}^{(1)}_{n,i} = \left(s_{i}^{\mathrm{e}}\mathbf{w}^H{\bf{G}}{\rm{diag}}({{\bf{h}}_n}) + \mathbf{e}_{n,i}\right)^{H}$}

				\State { Estimate $\hat{\mathbf{x}}$ from $({y}_{n,i},\mathbf{\tilde{u}}^{(1)}_{n,i})$  according to distributed Algorithm \ref{alg:algorithm_SDTLS} in Section \ref{section3C}.}
				
				\Statex {\!\!\!\!\!\!\!\!\!\!\textbf{Step 2:}}
				
				\State {$ \hat{y}^{(1)}_{n,i} = y_{n,i} - s_{i}^{\mathrm{e}}\mathbf{w}^H{\bf{G}}{\rm{diag}}({{\bf{h}}_n})\hat{\mathbf{x}}$}
				
				\State { Perform SVD on $\hat{\mathbf{Y}}_{n}^{(1)}=\hat{\mathbf{y}}_n^{(1)} \mathbf{z}_{n}^{H}$ to estimate data  $\hat{\mathbf{s}}^{\mathrm{d}} = \mathbf{Q}(:,1)$}
				
				\State{ $\mathbf{\tilde{u}}^{(2)}_{n,i}=\left({s}^{\mathrm{e}}_i\mathbf{w}^H{\bf{G}}{\rm{diag}}({{\bf{h}}_n})+\boldsymbol{\varpi}_n\right)^{H}$}
				
				\State { $\tilde{y}_{n,i}={y}_{n,i}-\hat{\mathbf{s}}^{\mathrm{d}}_i\mathbf{f}^{H}{\bf{G}}{\rm{diag}}({{\bf{h}}_n})\hat{\mathbf{x}}$}
				
				\State { Re-estimate $\hat{\mathbf{x}}$ from $(\tilde{y}_{n,i},\mathbf{\tilde{u}}^{(2)}_{n,i})$  according to distributed Algorithm \ref{alg:algorithm_SDTLS} in Section \ref{section3C}.}
				
			\end{algorithmic}
		\end{algorithm}
		
		{\bf{Remark 1}}:
		This work assumes that the sensing user network is only interested in targeted environment object sensing rather than communication data. Therefore, we discard the estimated data symbols after Algorithm \ref{alg:algorithm_SDTLS_two_step}. If the sensing users are interested in the data symbols, such data symbols can be stored.
		
		\subsection{Centralized Sensing Strategy}
		To estimate the ROI vector from the received signal, we need to design a specific sensing algorithm as mentioned in the proposed two-step scheme. First, we consider a centralized sensing strategy where we select a user with strong computing capabilities as the leader by assuming that the global estimated data is available to the leader user. For the centralized design, the leader user performs signal fusion and target sensing jointly by receiving signals from other users. Since we focus our analysis on the global data at the leader user without distinguishing the originality of the data, index $n$ is omitted in this section.
		Specifically, for any user, we have $y_{i}=\tilde{\mathbf{u}}^{H}_{i} \mathbf{x}^{0}$, where
		${\bf{\tilde{u}}}_{i}$ equals to ${\bf{\tilde{u}}}_{n,i} ^{(1)}$ or ${\bf{\tilde{u}}}_{n,i} ^{(2)}$ in the first or second step obtained in Algorithm \ref{alg:algorithm_SDTLS_two_step} by omitting index $n$, respectively, and $\mathbf{x}^0$ is the value of the true ROI.

		In general, the number of pixels occupied by a target is finite and often much smaller than the total number of pixels, i.e., elements of the vector $\mathbf{x}$ should be sparse.
		In practical scenarios, the target, such as a desk or a box, tends to exist in a continuous block rather than various scattered points in space \cite{shao}. In other words, the imaging vector $\mathbf{x}$ should have both element sparsity and block-wise sparsity characteristics. Motivated by scenarios where the unknown ROI coefficient appears in blocks rather than being arbitrarily spread spatially \cite{Sparse_LMS}, \cite{l0_LMS}, we propose to employ the penalty method to jointly exploit the element sparsity and structured sparsity of the ROI for recovering vector $\mathbf{x}$.

		Since the noise terms impair both the input and output data in \eqref{sens} and \eqref{see3}, we consider applying the total least-squares (TLS) method to reduce the impact of noise.
		The TLS method outperforms the least-squares (LS) based method when there are noisy input and output data because it can minimize the perturbations in the input and output data \cite{bertrand2012low}, \cite{DTLS-Li}.
		Considering the two-step sensing algorithm containing the interference from the data symbols or the impact of input noise, we propose the TLS-based two-step sensing algorithm.

		For brevity, we define $\left\| \mathbf{x}\right\|_{2,1} = \sum_{i=1}^{K_{2} \times K_{3}}\left\| \mathbf{x}^{i} \right\|_{2} $ as the $l_{2,1}$-norm of $\mathbf{x}$, where $\mathbf{x}$ is partitioned into $K_{2}\times K_{3}$ number of vectors, i.e., $\mathbf{x}=\operatorname{col}\left\{\mathbf{x}^{1}, \ldots,\mathbf{x}^{i}, \dots, \mathbf{x}^{K_{2}\times K_{3}}\right\}$ with $\mathbf{x}^{i}\in \mathbb{C}^{K_{1}}$.
		Then, the objective of the TLS method is to minimize the perturbations in both the input and output data, taking the element sparsity and the structured sparsity into account, which can be formulated as
		\begin{equation}
			\begin{aligned}
				\label{global_cost}
				J^{\mathrm{global}}(\mathbf{x} )\!&=\! \frac{E\left[\left\|\tilde{y}_{i}-\mathbf{\tilde{u}}_{i}^{H} \mathbf{x}\right\|_{2}^{2}\right]}{\|\mathbf{x}\|_{2}^{2}+1}
				\!+ \!{\eta}_{1}\!\sum_{k=1}^{K}|{x}_{k}| \!+\!{\eta}_{2}\! \sum_{i=1}^{K_{2} \times K_{3}}\!\left\| \mathbf{x}^{i} \right\|_{2}\\
				&= F(\mathbf{x})+ {\eta}_{1}\| \mathbf{x}\|_{1} +{\eta}_{2}\|\mathbf{x}\|_{2,1},
			\end{aligned}
		\end{equation}
		where $\eta_{1}\geq 0$ and $\eta_{2} \geq 0$ are the regularization parameters adopted to control the intensity of the element-sparsity penalty and block-wise sparsity penalty, which are expressed by $l_{1}$-norm and $l_{2,1}$-norm of $\mathbf{x}$, respectively.
		
		Note that the $l_1$-norm regularization term in equation \eqref{global_cost} encourages the ROI vectors to have few non-zero elements, while the $l_{2,1}$-norm regularization term promotes the continuity of these non-zero elements. This dual regularization approach results in both element-wise and structural sparsity in the estimated ROI vector, eliminating the requirement for prior knowledge of the number of empty pixels.

		In order to accurately estimate $\mathbf{x}$, we exploit the gradient descent method to search for the minimum of (\ref{global_cost}), which can be formulated as $\mathbf{x}_{i}=\mathbf{x}_{i-1}-\mu \Delta \mathbf{x}_{i}$.
		Herein, $\mathbf{x}_{i} $ is the global estimate obtained at time $i$, $\mu \ge 0$ is the updated step-size, and $\Delta \mathbf{x}_{i}$ is the derivative of $J^{\operatorname{global}}(\mathbf{x})$  with respect to $\mathbf{x}$ at time $i$.
		Next, we describe the gradient of each term in (\ref{global_cost}). Specifically, we denote $f(\mathbf{x}) =\frac{\partial F(\mathbf{x})}{\partial \mathbf{x}}$ as the gradient of $F(\mathbf{x})$ with respect to $\mathbf{x}$, which is expressed as
		\begin{equation}
		\begin{aligned}
			\label{gradient_for_F}
			f(\mathbf{x}) =-2\left(\frac{E\left[\tilde{\mathbf{u}}_{i}\left(\tilde{y}_{i}-\tilde{\mathbf{u}}_{i}^{H} \mathbf{x}\right)\right]}{\|\mathbf{x}\|_{2}^{2}+1}+\frac{E\left[\left(\tilde{y}_{i}-\tilde{\mathbf{u}}_{i}^{H} \mathbf{x}\right)^{2} \mathbf{x}\right]}{\left(\|\mathbf{x}\|_{2}^{2}+1\right)^{2}}\right).
		\end{aligned}
		\end{equation}
		
		However, since the signal is transmitted in the form of a data stream and sensing symbols arrive at devices one by one, it is difficult to obtain expectation values in (\ref{gradient_for_F}) during transmission of sensing symbols.
		As a result, we replace expectation values in above derivative with local instantaneous approximations \cite{lopez2014distributed}, \cite{DLMS}, i.e., $E\left[\tilde{y}_{i} \tilde{\mathbf{u}}_{i}\right] $ is replaced with $\tilde{y}_{i} \tilde{\mathbf{u}}_{i}$, while $E\left[\tilde{\mathbf{u}}_{i} \tilde{\mathbf{u}}_{i}^{H}\right]$ is replaced with $\tilde{\mathbf{u}}_{i} \tilde{\mathbf{u}}_{i}^{H}$ that yield the following instantaneous gradient
		\begin{align}
			\label{centralized_replaced}
			\hat{f}(\mathbf{x}) \!=\!-2 \frac{\tilde{y}_{i}-\tilde{\mathbf{u}}^{H}_{i} \mathbf{x}}{\|\mathbf{x}\|_{2}^{2}+1}\!\left(\!\tilde{\mathbf{u}}_{i}\!+\!\frac{\tilde{y}_{i}-\tilde{\mathbf{u}}_{i}^{H} \mathbf{x}}{\|\mathbf{x}\|_{2}^{2}+1} \mathbf{x}\!\right)
			\!=\!-2 \varepsilon_{i}\!\left(\tilde{\mathbf{u}}_{i}\!+\!\varepsilon_{i} \mathbf{x}\right),
		\end{align}
		where $\varepsilon_{i}=\frac{\tilde{y}_{i}-\mathbf{\tilde{u}}_{i}^{H} \mathbf{x}_{i-1}}{\left\|\mathbf{x}_{i-1}\right\|_{2}^{2}+1}$ is the weighted error at time $i$. The error due to the proposed approximation and the convergence will be detailed in the performance analysis of Section \ref{subsection:performance_analysis}.
		
		Next, by calculating the derivative of $J^{\operatorname{global}}(\mathbf{x})$  with respect to $\mathbf{x}$, we have
		\begin{equation}
			\begin{aligned}
				\Delta \mathbf{x}_{i} = \hat{f}(\mathbf{x}_{i-1})  +  \eta_{1}\operatorname{sign}\left(\mathbf{x}_{i-1}\right) +\eta_{2}\mathbf{\Sigma}(\mathbf{x}_{i-1}) \mathbf{x}_{i-1} ,
			\end{aligned}
		\end{equation}
		where $\operatorname{sign}(\mathbf{x}) = \left[\operatorname{sign}(x_{1}),\dots, \operatorname{sign}(x_{K}) \right]^{H}$ is the derivative of $\|\mathbf{x}\|_{1}$ with respect to $\mathbf{x}$, $\mathbf{\Sigma}(\mathbf{x}) \mathbf{x}$ is the derivative of $\|\mathbf{x}\|_{2,1}$ with respect to $\mathbf{x}$.
		Herein, $\mathbf{\Sigma}(\mathbf{x})$ is a diagonal matrix, i.e.,  $\mathbf{\Sigma}(\mathbf{x}) = {\rm{diag}}\{g(\mathbf{x}^{1})\mathbf{I}_{K_{1}} ,g(\mathbf{x}^{2})\mathbf{I}_{K_{1}} ,\dots,g(\mathbf{x}^{K_{2}\times K_{3}})\mathbf{I}_{K_{1}}   \} $, with
		$
		g(\mathbf{x}^{i}) =  \begin{cases} \frac{1}{\left\|\mathbf{x}^{i}\right\|_{2}} ,  & \left\|\mathbf{x}^{i}\right\|_{2} \neq 0 \\0, & \left\|\mathbf{x}^{i}\right\|_{2} = 0 \end{cases}
		$, and the sign function is defined as
		$
		\operatorname{sign}\left(x_{k}\right)= \begin{cases}\frac{x_{k}}{\left|x_{k}\right|}, & x_{k} \neq 0 \\ 0, & \text { else }\end{cases}
		$.
		Thus, we obtain an iterative update strategy for centralized sensing:
		\begin{equation}
		\begin{aligned}			\label{centralized_TLS_update}
			\mathbf{x}_{i}=\mathbf{x}_{i-1}+\mu \left[ \varepsilon_{i}\left(\tilde{\mathbf{u}}_{i}+\varepsilon_{i} \mathbf{x}_{i-1}\right) \right. \qquad \qquad \\
			\left. \qquad \qquad - \eta_{1}\operatorname{sign}\left(\mathbf{x}_{i-1}\right)  -\eta_{2} \mathbf{\Sigma}(\mathbf{x}_{i-1}) \mathbf{x}_{i-1} \right].
		\end{aligned}
		\end{equation}
		After multiple iterations of updates, the leader user will achieve a stable estimate of $\mathbf{x}^0$. However, the other non-leader users do not obtain sensing results since the global sensing results are only calculated by the leader user.
		
		\subsection{Distributed Sensing Strategy}
		\label{section3C}
		Centralized sensing requires all users to convey all their data to the leader user, which generates high communication signaling overhead for the leader user. Furthermore, relying on a single leader node can create a system performance bottleneck due to its potential malfunction. 
		To address these issues, we aim to design a distributed sensing strategy for the considered ISAC system. In particular, each user in turn improves the local ROI estimation accuracy by fusing the received contributions, obviating the need for a centralized data center or a leader node. Similar to the proposed centralized sensing strategy, we consider applying TLS estimation methods to reduce the impacts caused by input and output noises.
		
		In this paper, the weighted mean square error (MSE) between the estimated target information and the actual target information is considered a measure of sensing accuracy.
		By assuming that ${\bf{\tilde{u}}}_{n,i}$ is ${\bf{\tilde{u}}}_{n,i} ^{(1)}$ or ${\bf{\tilde{u}}}_{n,i} ^{(2)}$ as illustrated in the first or second step of Algorithm \ref{alg:algorithm_SDTLS_two_step}, respectively, the local cost function at user $n$ is defined as the local weighted MSE combined with the $l_{1}$ and $l_{2,1}$ norm, which is given by
		\begin{equation}
		\begin{aligned}
			J_{n}^{\mathrm{loc}}\!(\mathbf{x} )\!&=\!\sum_{\!l \in N_{n}\!}\! c_{l, n}\!\frac{\!E\!\left[\!\left\|\tilde{y}_{l, i}\!-\!\mathbf{\tilde{u}}_{l, i}^{H} \mathbf{x}\right\|_{2}^{2}\!\right]\!}{\|\mathbf{x}\|_{2}^{2}+1}\!+\!{\eta}_{1}\!\sum_{k=1}^{K}\!|{x}_{k}|\!+\!{\eta}_{2}\!\! \sum_{i=1}^{\!K_{2}\! \times\! K_{3}\!}\!\left\|\! \mathbf{x}^{i}\! \right\|_{2} \\
			&=\sum_{l \in N_{n}} c_{l, n} F_{l}(\mathbf{x})+ {\eta}_{1}\| \mathbf{x}\|_{1}+{\eta}_{2}\|\mathbf{x}\|_{2,1}
			\label{DTLS_loc_functon1},
		\end{aligned}
		\end{equation}
		where $F_{l}(\mathbf{x})=\frac{E\left[\left\|\tilde{y}_{l, i}-\mathbf{\tilde{u}}^{H}_{l, i} \mathbf{x}\right\|_{2}^{2}\right]}{\|\mathbf{x}\|_{2}^{2}+1}$, $N_{n}$ is the neighbor set of user $n$, and $\mathbf{C} = \{ c_{ln} \}$ is defined as an $N\times N$  non-negative cooperative coefficient matrix, which satisfies $\mathbf{1}^{T} \mathbf{C}=\mathbf{1}^{T}$,   $\mathbf{C} \mathbf{1} = \mathbf{1}$, and $c_{l, n}=0 \text { if } l \notin N_{n}$. Herein, $\mathbf{1}$ is an $N \times 1$  unity vector. Since $\sum_{l \in N_{n}} c_{l, n}=1$, (\ref{DTLS_loc_functon1}) can be rewritten as
		\begin{align}
			\label{DTLS_loc_function2}
			J_{n}^{\mathrm{loc}}(\mathbf{x})=\sum_{l \in N_{n}} c_{l, n}\left(F_{l}(\mathbf{x}) + \eta_{1}\|\mathbf{x}\|_{1} +\eta_{2}\|\mathbf{x}\|_{2,1} \right).
		\end{align}
		
		\begin{algorithm}[t]
			\caption{Distributed Sensing Algorithm}  \label{alg:algorithm_SDTLS}
			\textbf{Inputs:}
			$\tilde{y}_{n,i},\mathbf{\tilde{u}}_{n,i}, \forall n\leq N, \forall i \leq T$
			
			\textbf{Initialize:} Initialize $\mathbf{x}_{n,0}$ for each user $n$, step size $\mu_{n}$, regularization parameter $\eta_{1}$, $\eta_{2}$, and cooperative coefficient matrix $\mathbf{C}$.
			\begin{algorithmic}[1]
				\For{$i = 1:T$}
				\For{$n = 1:N$}
				\State  	\textbf{Adaptation:}
				\State	$\quad \varepsilon_{n, i}=\frac{\tilde{y}_{n, i}-\mathbf{\tilde{u}}_{n, i}^{H} \mathbf{x}_{n, i-1}}{\left\|\mathbf{x}_{n, i-1}\right\|_{2}^{2}+1}$
				\State $\quad \boldsymbol{\varphi}_{n, i}=\mathbf{x}_{n, i-1}+\mu_{n}\left(\varepsilon_{n, i}\left(\mathbf{\tilde{u}}_{n, i}+\varepsilon_{n, i} \mathbf{x}_{n, i-1}\right) \right.$
				\Statex $\qquad \qquad \qquad \quad	\left.-\eta_{1} \operatorname{sign}(\mathbf{x}_{n,i-1})\!-\!\eta_{2} \mathbf{\Sigma}(\mathbf{x}_{n,i-1}) \mathbf{x}_{n, i-1}   \right)$
				\State \textbf{Communication:}
				\State $\quad $Transmit $\boldsymbol{\varphi}_{n,i}$ to all the neighbors in $N_{n}$
				\State \textbf{Combination:}
				\State	$\quad \mathbf{x}_{n, i}=\sum_{l \in N_{n}} c_{l, n} \boldsymbol{\varphi}_{l, i}$
				\EndFor
				\EndFor
			\end{algorithmic}
		\end{algorithm}
		
		Similar to (\ref{centralized_replaced}), we replace the expectation values with instantaneous approximations when calculating the gradient of $F_l(\mathbf{x})$ with respect to $\mathbf{x}$.
		We define the error caused by this approximation as the gradient error, which will be analyzed in Section \ref{subsection:performance_analysis}.
		Additionally, similar to the centralized update in (\ref{centralized_TLS_update}), we can easily obtain an updated formula for the ROI vector $\mathbf{x}$ of each user.
		Note that when we calculate the update weights for each user, the observation data from all neighboring users of $n$ needs to be transferred to user $n$.
		In practice, conveying the raw data of all users will introduce a considerable communication burden to the whole network.
		To solve this problem, we calculate an intermediate estimate for each user at each moment and convey it to neighboring users to reduce the burden.
		Specifically, we divide the distributed update into two steps with adaptation and combination.
		First, each user calculates the instantaneous gradient in the adaptation step and derives an intermediate estimate for sending to neighboring users.
		Then, in the combination step, each user fuses intermediate estimates sent from neighboring users and adaptively updates itself.
		By implementing these two steps alternately, users communicate and collaborate with each other to minimize global loss and jointly estimate optimal ROI.
		Based on these considerations, we propose a distributed update strategy derived as
		\begin{align}
			\label{DTLS_solved}
			\left\{\begin{array}{l}
				\boldsymbol{\varphi}_{n, i}=\mathbf{x}_{n, i-1}+\mu_{n}\left(\varepsilon_{n, i}\left(\mathbf{\tilde{u}}_{n, i}+\varepsilon_{, i} \mathbf{x}_{n, i-1}\right)\right.\\
				\left.\qquad \quad-\eta_{1} \operatorname{sign}(\mathbf{x}_{n,i-1}) -\eta_{2} \mathbf{\Sigma}(\mathbf{x}_{n,i-1}) \mathbf{x}_{n, i-1}   \right), \\
				\mathbf{x}_{n, i}=\sum_{l \in N_{n}} c_{l, n} \boldsymbol{\varphi}_{l, i},
			\end{array}\right.
		\end{align}
		where $\boldsymbol{\varphi}_{n, i}$ and $\varepsilon_{n,i}=\frac{\tilde{y}_{n,i}-\mathbf{\tilde{u}}_{n,i}^{H} \mathbf{x}_{n,i-1}}{\left\|\mathbf{x}_{n,i-1}\right\|_{2}^{2}+1}$ are the intermediate estimate and the weighted error for user $n$ at time $i$, respectively.
		Notice that this update scheme follows the adapt-then-combine (ATC) scheme in the distributed estimation \cite{9141196}. 
		
\begin{figure}[t!]
			\setlength{\abovecaptionskip}{-0.cm}
			\setlength{\belowcaptionskip}{0.cm}
			\centering
			\includegraphics [width=0.45\textwidth] {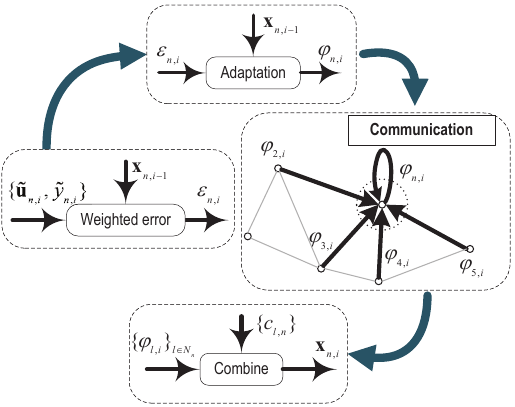}
			\caption{Illustration of the distributed sensing scheme for node $n$ at time $i$ in Algorithm \ref{alg:algorithm_SDTLS}.}
			\label{Distributed_Sensing_scheme}
		\end{figure}

		The details of the parameters estimation are specified in Algorithm \ref{alg:algorithm_SDTLS}. 		
		In general, the $n$-th user performs four steps at the $i$-th iteration.
		In the first step (line 4-5), each user uses its input value $({y}_{n,i},\mathbf{\tilde{u}}_{n,i})$ available at the $i$-th iteration to adaptively adjust its weighted error, i.e., $\varepsilon_{n,i}$, and the intermediate estimate $\boldsymbol{\varphi}_{n, i}$.
		In the second step (line 7), each node transmits its intermediate estimate $\boldsymbol{\varphi}_{n, i}$ forward to its neighboring nodes.
		In the third step (line 9), each node combines the obtained intermediate estimates to obtain an estimate $\mathbf{x}_{n, i}$	 for the current iteration.
		In each iteration of the algorithm, the intermediate estimates are transmitted among different users and gradually converge to the vicinity of true ROI with the iterative implementation of the algorithm.
		Fig. \ref{Distributed_Sensing_scheme} shows schematically the distributed cooperation scheme for $n$-th user at $i$-time.

		Since $K$ parameters are needed to be computed during the calculation for the approximated gradient, computing the intermediate estimate requires $\mathcal{O}(K)$ operations in each iteration for each user $n$.
		As for the combination step, for each user $n$, it needs $\mathcal{O}(|N_n|K)$ number of operations to merge the intermediate estimates sent by the neighbors, where $|N_n|$ is the cardinality of $N_n$.
		Therefore, for total users $N$ and the $T$ iterations, the computational complexity of Algorithm \ref{alg:algorithm_SDTLS} can be derived as $\mathcal{O}(KT+KT\Sigma_{n=1}^{N}|N_n|)$.
		In addition, the computational complexity of Algorithm \ref{alg:algorithm_SDTLS_two_step} is expressed as $\mathcal{O}(T^2 + 2(KT + KT\Sigma_{n=1}^{N}|N_n|))$, where $\mathcal{O}(T^2)$ accounts for the computational complexity of SVD applied to $\hat{\mathbf{Y}}_{n}^{(1)}$.
		For communication overhead, in the collaboration step, each user $n$ sends $K$ intermediate parameters to its $|N_n|$ neighbors in each iteration.
		Thus, the communication cost of Algorithm \ref{alg:algorithm_SDTLS} in the whole network is given by $\mathcal{O}(KT\Sigma_{n=1}^{N}|N_n|)$.
		Comparatively, in the centralized sensing strategy, nodes only transmit input vectors to the leader node, so the communication overhead of the centralized algorithm is $\mathcal{O}(KTN)$.

		Although the centralized scheme has lower communication overhead than the proposed distributed scheme, its robustness is compromised due to its heavy reliance on a leader node with high computing power.

		Through a distributed update scheme with multi-device collaboration, each user can eventually obtain a reliable estimate of the ROI. In the following subsection, we will provide an elaborated proof for the convergence of the distributed update strategy along with the corresponding sensing performance metrics.

		\subsection{Performance Metrics for Distributed Sensing}
		\label{subsection:performance_analysis}
		In order to verify the convergence properties of the proposed algorithm and jointly optimize sensing and communication performance, we will conduct a performance analysis and derive corresponding performance metrics for distributed sensing in this section.

		The theoretical deviation of the sensed ROI from the true value is mainly caused by replacing the expectation with an instantaneous approximation. Therefore, we adopt gradient error to model this deviation, which is defined as
		$\mathbf{m}_{l}\left(\mathbf{x}_{l,i-1}\right)=\hat{f}_{l}\left(\mathbf{x}_{l,i-1}\right)-f_{l}\left(\mathbf{x}_{l,i-1}\right)$. Herein, $f_{l}(\mathbf{x})$ is the gradient of $F_{l}(\mathbf{x})$ with respect to $\mathbf{x}$, which is given by
		\begin{equation}
		\begin{aligned} f_{l}(\mathbf{x})&=\frac{\partial F_{l}(\mathbf{x})}{\partial \mathbf{x}} \\ &=\!-2\!\left(\!\frac{\!E\!\left[\tilde{\mathbf{u}}_{l, i}\!\left(\tilde{y}_{l, i}\!-\!\tilde{\mathbf{u}}_{l, i}^{H} \mathbf{x}\right)\!\right]}{\|\mathbf{x}\|_{2}^{2}+1}\!+\!\frac{\!E\!\left[\!\left(\tilde{y}_{l,i}\!-\!\tilde{\mathbf{u}}_{l, i}^{H} \mathbf{x}\right)\!^{2}\! \mathbf{x}\right]\!}{\left(\|\mathbf{x}\|_{2}^{2}+1\right)^{2}}\!\right)\!,
		\end{aligned}
		\end{equation}
		and $\hat{f}_{l}(\mathbf{x})$ is the instantaneous approximation for $f_{l}(\mathbf{x})$ by replacing the expectation values with the local instantaneous approximations, namely, $E\left[\tilde{y}_{l, i} \tilde{\mathbf{u}}_{l, i}\right] \approx \tilde{y}_{l, i} \tilde{\mathbf{u}}_{l, i}$, and $E\left[\tilde{\mathbf{u}}_{l, i} \tilde{\mathbf{u}}_{l, i}^{H}\right] \approx \tilde{\mathbf{u}}_{l, i} \tilde{\mathbf{u}}_{l, i}^{H}$. In this context, $\hat{f}_{l}(\mathbf{x})$ can be expressed as
		\begin{equation}
		\begin{aligned}
			\hat{f}_{l}(\mathbf{x}) &=-2 \frac{\tilde{y}_{l, i}-\tilde{\mathbf{u}}_{l, i}^{H} \mathbf{x}}{\|\mathbf{x}\|_{2}^{2}+1}\left(\tilde{\mathbf{u}}_{l, i}+\frac{\tilde{y}_{l, i}-\tilde{\mathbf{u}}_{l, i}^{H} \mathbf{x}}{\|\mathbf{x}\|_{2}^{2}+1} \mathbf{x}\right)\\
			&=-2 \varepsilon_{l, i}\left(\tilde{\mathbf{u}}_{l, i}+\varepsilon_{l, i} \mathbf{x}\right),
		\end{aligned}
		\end{equation}
		where $\varepsilon_{l, i}=\frac{\tilde{y}_{l, i}-\mathbf{\tilde{u}}_{l, i}^{H} \mathbf{x}_{l, i-1}}{\left\|\mathbf{x}_{l, i-1}\right\|_{2}^{2}+1}$ is the weighted error for user $l$ at time $i$.
		
		Then, we can obtain the Hessian matrix of $F_{l}(\mathbf{x})$ by computing the second-order derivative with respect to $\mathbf{x}$
		\begin{align}
			H_{l}(\mathbf{x})= \frac{2}{\|\mathbf{x}\|_{2}^{2}+1}
			\left(\tilde{\mathbf{R}}_{l}-f_{l}(\mathbf{x}) \mathbf{x}^{H}-\mathbf{x} f_{l}^{H}(\mathbf{x})-F_{l}(\mathbf{x}) \mathbf{I}_{l}\right),
		\end{align}
		where $\tilde{\mathbf{R}}_{l}=E\left[\tilde{\mathbf{u}}_{l, i} \tilde{\mathbf{u}}_{l, i}^{H}\right]$.
		It is assumed that the estimate converges to the vicinity of $\mathbf{x}^{0}$, as commonly adopted in the analysis of distributed estimate algorithms \cite{DTLS-Li}, \cite{TLS-Arab}.
		Then, we obtain
		\begin{align}
			H_{l}\left(\mathbf{x}^{0}\right) =\frac{2 \mathbf{R}_{l}}{\left\|\mathbf{x}^{0}\right\|_{2}^{2}+1},
		\end{align}
		where
		\begin{align}\label{R_l}
			\mathbf{R}_{l}=E\left[\mathbf{u}_{l, i} \mathbf{u}_{l, i}^{H}\right],\end{align}
		is the covariance matrix of $\mathbf{u}_{l, i}$ with
		\begin{align}\label{U_li}
			{\bf{u}}_{l,i} = \left( {s}_{i}^{\mathrm{e}}\mathbf{w}^H{\bf{G}}{\operatorname{diag}}({{\bf{h}}_l})\right)^{H},\end{align}
		being the input signal without noise. Then, the global Hessian matrix can be defined as $\mathbf{H} = \operatorname{diag}\{ {H}_{1}(\mathbf{x}^{0}),\dots,{H}_{N}(\mathbf{x}^{0})\}$.
		
		To obtain the global mean square error (MSE), we define $\mathbf{x}_{i}=\operatorname{col}\left\{\mathbf{x}_{1, i}, \ldots, \mathbf{x}_{K, i}\right\}$ and $\mathbf{x}^{(0)}=\operatorname{col}\left\{\mathbf{x}^{0}, \ldots, \mathbf{x}^{0}\right\}$.
		For the global error vector $\overline{\mathbf{x}}_{i}=\mathbf{x}^{0}-\mathbf{x}_{i}$, we can evaluate its weighted norm as follows
		\begin{equation}
			\label{weight_norm_error1}
			\begin{aligned}
				E {\left[\left\|\overline{\mathbf{x}}_{i}\right\|_{\mathbf{\Lambda}}^{2}\right] }
				=& E\left[\left\|\overline{\mathbf{x}}_{i-1}\right\|_{\mathbf{\Phi}}^{2}\right]+E\left[\|\mathbf{m}\|_{\mathbf{T}}^{2}\right]\\
				&+ E\left[\left\|\eta_{1} \operatorname{sign}(\mathbf{x}_{i-1})\right\|_{\Lambda^{\prime}}^{2}+ \eta_{2}\mathbf{\Sigma}(\mathbf{x}_{i-1}) \mathbf{x}_{i-1}\right] \\
				&+2  E\left[\left(\eta_{1}\operatorname{sign}(\mathbf{x}_{i-1})   + \eta_{2}\mathbf{\Sigma}(\mathbf{x}_{i-1}) \mathbf{x}_{i-1}\right) \right. \\
				&\left. \quad \times\mathbf{D} \mathbf{O}^{H} \mathbf{\Lambda} \mathbf{O}\left(\mathbf{I}_{KN}-\mathbf{D} \mathbf{H}\right) \overline{\mathbf{x}}_{i-1}\right],
			\end{aligned}
		\end{equation}
		where the weighting matrix $\mathbf{\Lambda}$ is any Hermitian semipositive-definite matrix to be selected freely, $\mathbf{\Lambda}^{\prime}=\mathbf{D} \mathbf{O}^{H} \mathbf{\Lambda} \mathbf{O} \mathbf{D}$, $\mathbf{D}=\operatorname{diag}\left\{\mu_{1} \mathbf{I}_{K}, \ldots, \mu_{N} \mathbf{I}_{K}\right\}$, $\mathbf{O}=\mathbf{C}^{H} \otimes \mathbf{I}_{K}$, and $\mathbf{m}=\operatorname{col}\left\{\mathbf{m}_{1}\left(\mathbf{x}^{0}\right), \ldots, \mathbf{m}_{N}\left(\mathbf{x}^{0}\right)\right\}$.
		$\left\|\overline{\mathbf{x}}_{i}\right\|_{\mathbf{\Lambda}}^{2} = \overline{\mathbf{x}}_{i}^{H} \mathbf{\Lambda} \overline{\mathbf{x}}_{i}$ is defined as the weighted Euclidean norm of the vector $\overline{\mathbf{x}}_{i-1}$ with $\mathbf{\Lambda}$ being the weight matrix.
		In addition, $\mathbf{\Phi}$ and $\mathbf{T}$ are the weighted matrices defined as
		\begin{align}
			\mathbf{\Phi}=\left(\mathbf{I}_{K N}-\mathbf{D} \mathbf{H}\right) \mathbf{O}^{H} \mathbf{\Lambda} \mathbf{O}\left(\mathbf{I}_{K N}-\mathbf{D} \mathbf{H}\right),
		\end{align}
		and
		\begin{align}
			\mathbf{T} = \mathbf{D} \mathbf{O}^{H}\mathbf{\Phi}\mathbf{O}\mathbf{D}.
		\end{align}
		
		The matrices defined above can be expressed as vectors to facilitate the subsequent calculations, i.e., $\boldsymbol{\lambda}=\operatorname{vec}\{\boldsymbol{\Lambda}\}$, $\boldsymbol{\phi}=\operatorname{vec}\{\boldsymbol{\Phi}\}$, $\mathbf{t}=\operatorname{vec}\{\mathbf{T}\}$.
		Consequently, we have $\boldsymbol{\phi}=\mathbf{P} \boldsymbol{\lambda}$, $\mathbf{t}=\mathbf{V} \boldsymbol{\lambda}$ and $E\left[\|\mathbf{m}\|_{\mathbf{T}}^{2}\right] =\mathbf{q}^{H} \mathbf{t}$, where
		\begin{align}\label{p0}
			\mathbf{P}&=\left(\mathbf{I}_{K N}-\mathbf{D} \mathbf{H}\right) \mathbf{O}^{H} \otimes\left(\mathbf{I}_{K N}-\mathbf{D} \mathbf{H}\right) \mathbf{O}^{H},\\
			\mathbf{V}&=\mathbf{D} \mathbf{O}^{H} \otimes \mathbf{D} \mathbf{O}^{H},\\
			\label{q0}
			\mathbf{q} &= \operatorname{vec}\{\mathbf{Q}\},
		\end{align}
		with
		\begin{align}\label{Q}
			\mathbf{Q}\!=\! \operatorname{diag}\{E\!\left[\mathbf{m}_{1}\!\left(\mathbf{x}^{0}\right)\! \mathbf{m}_{1}^{H}\!\left(\mathbf{x}^{0}\right)\!\right],\dots, \!E\!\left[\mathbf{m}_{N}\!\left(\mathbf{x}^{0}\right)\! \mathbf{m}_{N}^{H}\!\left(\mathbf{x}^{0}\right)\!\right]\! \}.
		\end{align}
		Herein, for $l=1,\dots,N$, we have
		\begin{equation}
		\begin{aligned}
			&E\left[\mathbf{m}_{l}\left(\mathbf{x}^{0}\right) \mathbf{m}_{l}^{H}\left(\mathbf{x}^{0}\right)\right]\\
			& \qquad  \qquad=\frac{4 \boldsymbol{\sigma}_{\text { in }}^{2}}{\left\|\mathbf{x}^{0}\right\|^{2}+1} \left(\mathbf{R}_{l}+\boldsymbol{\sigma}_{\text{ in }}^{2} \mathbf{I}_{l}
			-\frac{3 \boldsymbol{\sigma}_{\text { in }}^{2} \mathbf{x}^{0}\left(\mathbf{x}^{0}\right)^{H}}{\left\|\mathbf{x}^{0}\right\|^{2}+1}\right),
		\end{aligned}
		\end{equation}	
		where $\mathbf{R}_{l}=E\left[\mathbf{u}_{l, i} \mathbf{u}_{l, i}^{H}\right]$.
		Then, we can rewrite the weighted norm of the error vector in (\ref{weight_norm_error1}) as
		\begin{align}
			E\left[\left\|\overline{\mathbf{x}}_{i}\right\|_{\boldsymbol{\lambda}}^{2}\right]=E\left[\left\|\overline{\mathbf{x}}_{i-1}\right\|_{\mathbf{P} \boldsymbol{\lambda}}^{2}\right]+\mathbf{q}^{H} \mathbf{V} \boldsymbol{\lambda}+\boldsymbol{\chi}_{\boldsymbol{\Lambda}, i}(\eta_{1},\eta_{2}),
		\end{align}
		where
		\begin{equation}
			\begin{aligned}
				\boldsymbol{\chi}_{\boldsymbol{\Lambda}, i}(\eta_{1},\eta_{2}) =&  E\left[\left\|\eta_{1}\operatorname{sign}(\mathbf{x}_{i-1})\right\|_{\Lambda^{\prime}}^{2}   + \eta_{2} \mathbf{\Sigma}(\mathbf{x}_{i-1}) \mathbf{x}_{i-1} \right] \\
				&+2  E\left[\left(\eta_{1} \operatorname{sign}(\mathbf{x}_{i-1})   + \eta_{2} \mathbf{\Sigma}(\mathbf{x}_{i-1}) \mathbf{x}_{i-1} \right) \right.\\ 
				&\left. \quad \times\mathbf{D} \mathbf{O}^{T} \mathbf{\Lambda} \mathbf{O}\left(\mathbf{I}_{KN}-\mathbf{D} \mathbf{H}\right) \overline{\mathbf{x}}_{i-1}\right] .
			\end{aligned}
		\end{equation}
		Note that both $\mathbf{\Sigma}(\mathbf{x}_{i-1}) \mathbf{x}_{i-1}$ and $\operatorname{sign}(\mathbf{x}_{i-1})$ are bounded for any time $i$. Therefore, the term $\boldsymbol{\chi}_{\boldsymbol{\Lambda}, \infty}(\eta_{1},\eta_{2})=\lim\limits_{i\to+\infty}\boldsymbol{\chi}_{\boldsymbol{\Lambda}, i}(\eta_{1},\eta_{2})$ is upper-bounded. In other words, it means that as $i \rightarrow \infty$, we have
		\begin{align}
			E\left[\left\|\overline{\mathbf{x}}_{\infty}\right\|_{\boldsymbol{\lambda}}^{2}\right]=E\left[\left\|\overline{\mathbf{x}}_{\infty}\right\|_{\mathbf{P} \boldsymbol{\lambda}}^{2}\right]+\mathbf{q}^{H} \mathbf{V} \boldsymbol{\lambda}+\boldsymbol{\chi}_{\boldsymbol{\Lambda}, \infty}(\eta_{1},\eta_{2}).
		\end{align}
		
		In this context, if we set $\boldsymbol{\lambda}=\left(\mathbf{I}_{M^{2} N^{2}}-\mathbf{P}\right)^{-1} \operatorname{vec}\left\{\mathbf{I}_{M N}\right\}$ and $\boldsymbol{\Lambda}=\mathbf{I}$, we can obtain the steady-state MSE between the estimated ROI vector $\mathbf{x}$ and the actual vector $\mathbf{x}^{0}$ as
		\begin{equation}
		\begin{aligned}
			\label{steady-state-MSE}
			\text{MSE} &= E\left[\left\|\overline{\mathbf{x}}_{\infty}\right\|_{2}^{2}\right]\\
			&=\mathbf{q}^{H} \mathbf{V}\left(\mathbf{I}_{K^{2} N^{2}}-\mathbf{P}\right)^{-1} \operatorname{vec}\left\{\mathbf{I}_{K N}\right\}+\boldsymbol{\chi}_{\mathbf{I}, \infty}(\eta_{1},\eta_{2}).
		\end{aligned}
		\end{equation}
		In order to conveniently calculate the MSE, we assume that the step sizes for all users are the same, i.e., $\mu_1 = \mu_2 = \dots = \mu_N = \mu$. Then, we have
		\begin{equation} \label{35}
			\begin{aligned}
				\boldsymbol{\chi}_{\mathbf{I}, \infty}(\eta_{1},\eta_{2})=& \mu^{2} E\left[\left\|\eta_{1}\operatorname{sign}(\mathbf{x}_{\infty}) + \eta_{2} \mathbf{\Sigma}(\mathbf{x}_{\infty}) \mathbf{x}_{\infty}   \right\|_{\mathbf{O}^{H}\mathbf{O}}^{2}\right] \\
				&+2 \mu\left(1-\frac{2 \mu \sigma_{in}^{2}}{\left\|\mathbf{x}^{0}\right\|_{2}^{2}+1}\right)  E\left[\left( \eta_{1}\operatorname{sign}(\mathbf{x}_{\infty})  \right.\right. \\ 
				& \left. \left. \qquad\qquad\qquad +  \eta_{2} \mathbf{\Sigma}(\mathbf{x}_{\infty}) \mathbf{x}_{\infty}\right) \mathbf{O}^{H}  \mathbf{O} \overline{\mathbf{x}}_{\infty}\right],
			\end{aligned}
		\end{equation}
		which has a constant upper-bound expressed as $\alpha$.
		Thus, the steady-state MSE is obtained as
		\begin{equation}
		\begin{aligned}
			\label{steady-state-MSE_1}
			\text{MSE} &= E\left[\left\|\overline{\mathbf{x}}_{\infty}\right\|_{2}^{2}\right]\\ &=\mathbf{q}^{H} \mathbf{V}\left(\mathbf{I}_{K^{2} N^{2}}-\mathbf{P}\right)^{-1} \operatorname{vec}\left\{\mathbf{I}_{K N}\right\}+\alpha.
		\end{aligned}
		\end{equation}
		
		If the step-size ${\mu}$ in matrix $\mathbf{D}$ is chosen to be small enough to ensure mean convergence, it is established that the matrix $\mathbf{DH}$ is stable. In other words, all its eigenvalues lie within the unit circle.
		Thus, the eigenvalues of matrix $\mathbf{I}_{K N}-\mathbf{D} \mathbf{H}$ also lie within the unit circle.
		Based on Lemma 1 in \cite{DLMS}, we can affirm that $\left(\mathbf{I}_{K N}-\mathbf{D} \mathbf{H}\right)\mathbf{O}^{H}$ is also stable, provided that $\mathbf{O}$ is a left-stochastic matrix.
		Given the expression of the matrix $\mathbf{P}$ in \eqref{p0}, we infer that its eigenvalues are the squares of those of the matrix $\left(\mathbf{I}_{K N}-\mathbf{D} \mathbf{H}\right) \mathbf{O}^{H}$.
		Thus, the matrix $\mathbf{P}$ is also stable, affirming the invertibility of $\mathbf{I}_{K^{2} N^{2}}-\mathbf{P}$. This, in turn, implies the stability of \eqref{steady-state-MSE_1}. Therefore, we confirm the convergence of the steady-state MSE in the mean square sense.

		{\bf{Remark 2:}} Note that the matrix ${\mathbf{P}}$ in (\ref{steady-state-MSE_1}) contains the adjacency matrix ${\mathbf{C}}$ that is determined by the network structure. Thus, the derived networked MSE in (\ref{steady-state-MSE_1}) can characterize the effect of network topology on estimation performance.
		
		\section{Joint Beamforming Design of Sensing and Communication} \label{section_4}
		In 6G wireless networks, the sensing and communication beamforming vectors for ISAC should be jointly designed to improve ISAC performance. Based on the derived networked sensing performance metric, the purpose of the ISAC system is to minimize sensing error and maximize communication quality subject to maximum transmit power constraint.

		Here, we consider adopting the communication signal-to-interference-plus-noise ratio (SINR) at the user to evaluate communication quality and the networked MSE of the distributed algorithm to evaluate sensing capability. We denote the received SINR at the communication user as
		\begin{equation}
			\begin{aligned}
				\mathcal{F}_1 = \frac{|\mathbf{g}^H\mathbf{f}|^2}{\sigma^2+|\mathbf{g}^H\mathbf{w}|^2},
			\end{aligned}
		\end{equation}
		which is a common performance metric for measuring communication capability \cite{liu2018toward}, and high SINR is expected to obtain high-quality service for wireless communications. On the other hand, by ignoring the constant in \eqref{steady-state-MSE_1},
		we obtain the following networked MSE as the performance metric for evaluating the sensing capability:
		\begin{equation}\label{MSE_f2}
			\begin{aligned}\mathcal{F}_2 = \mathbf{q}^{H} \mathbf{V}\left(\mathbf{I}_{K^{2} N^{2}}-\mathbf{P}\right)^{-1} \operatorname{vec}\left\{\mathbf{I}_{K N}\right\}.
			\end{aligned}
		\end{equation}
		
		Note that the expression in \eqref{MSE_f2} is captured by $\mathbf{P}$ and $\mathbf{q}$. For ease of presentation, we replace ${\bf{u}}_{n,i}$ with ${s}_{i}^{\mathrm{e}}\mathbf{w}^H{\bf{G}}{\rm{diag}}({{\bf{h}}_n})$ in (\ref{U_li}) to rewrite $\mathbf{q}$ and $\mathbf{P}$ as
		\begin{align}
			\mathbf{q} = \alpha_{1} \left({s}^{e}\right)^{2}\left(\mathbf{U}\otimes \mathbf{U}\right)
			\operatorname{vec}\{\mathbf{I}_{N}\otimes \mathbf{w}\mathbf{w}^{H}\}
			+\operatorname{vec}\{\mathbf{B}\},
		\end{align}
		and
		\begin{equation}
			\label{P_distributed}
			\begin{aligned}
				\!\mathbf{P}\!&=\mathbf{O}^{H}\otimes \mathbf{O}^{H}\\
				&\;\;-\!\alpha_{2}\! \left(\!{s}^{e}\!\right)^{2}\! \left\{\! \left[\mathbf{U}\!\left(\mathbf{I}_{N}\!\otimes\!  \mathbf{w}\mathbf{w}^{H}\right)\!\mathbf{U}^{H}  \right]\! \otimes \!\mathbf{I}_{KN}\! \right\}\! \left[\!\left( \mathbf{D}\!\mathbf{O}^{H}  \right)\!\otimes\! \mathbf{O}^{H}  \right]\!\\
				&\;\;-\! \alpha_{2}\! \left(\!{s}^{e}\!\right)^{2}\! \left\{\! \mathbf{I}_{KN}\! \otimes\! \left[\mathbf{U}\!\left(\mathbf{I}_{N}\!\otimes\!  \mathbf{w}\mathbf{w}^{H} \right)\!\mathbf{U}^{H}  \right]\right\}\! \left[ \mathbf{O}^{H} \!\otimes  \!\left( \mathbf{D}\mathbf{O}^{H}  \right)  \right] \\
				&\;\;+\!\alpha_{2}^{2}\! \left(\!{s}^{e}\!\right)^{4}\! \left\{\!\left[\mathbf{U}\!\left(\mathbf{I}_{N}\!\otimes\!  \mathbf{w}\mathbf{w}^{H} \right)\!\mathbf{U}^{H}  \right] \!\otimes\! \left[\mathbf{U}\!\left(\mathbf{I}_{N}\!\otimes \! \mathbf{w}\mathbf{w}^{H} \right)\!\mathbf{U}^{H}  \right]\! \right\}\\ 
				&\quad \times \left(\mathbf{D}\otimes \mathbf{D} \right) \left(\mathbf{O}^{H}\otimes \mathbf{O}^{H} \right),
			\end{aligned}
		\end{equation}
		respectively, where $\mathbf{B} = \mathbf{I}_{N} \otimes \left( \boldsymbol{\sigma}_{\text{ in }}^{2} \mathbf{I}_{K}
		-\frac{3 \boldsymbol{\sigma}_{\text { in }}^{2} \mathbf{x}^{0}\left(\mathbf{x}^{0}\right)^{H}}{\left\|\mathbf{x}^{0}\right\|^{2}+1}\right)$ is independent of $\mathbf{w}$ and $\mathbf{f}$, $\alpha_{1} = \frac{4 \boldsymbol{\sigma}_{\text { in }}^{2}}{\left\|\mathbf{x}^{0}\right\|^{2}+1}$ and $\alpha_{2} =\frac{2}{\left\|\mathbf{x}^{0}\right\|_{2}^{2}+1} $ are constants,
		$\mathbf{O}=\mathbf{C}^{H} \otimes \mathbf{I}_{K}$ describes the connectivity between users in the network, $\mathbf{D}=\operatorname{diag}\left\{\mu_{1} \mathbf{I}_{K}, \ldots, \mu_{N} \mathbf{I}_{K}\right\}$ measures the step-size for updating at different users, and $\mathbf{U} = \operatorname{diag}\{ \operatorname{diag}\left(\mathbf{h}_{1}\right),\dots, \operatorname{diag}\left(\mathbf{h}_{N}\right) \}\left( \mathbf{I}_{N} \otimes \mathbf{G}^{H}\right)$ covers the channel state information from the BS to the ROI and from the ROI to different users.
		
		Obviously, the matrix $\mathbf{P}$ contains a large number of higher-order terms with respect to $\mathbf{O}$ and $\mathbf{D}$, which complicates the optimization of $\bf{f}$ and $\bf{w}$.
		Fortunately, since the values of each element in $\mathbf{O}$ and $\mathbf{D}$ are less than $1$, which is determined by both the elements in $\mathbf{C}$ and $\mathbf{\mu}_k, \forall k$, being less than 1, we can ignore the third and fourth-order terms with respect to $\mathbf{O}$ and $\mathbf{D}$ in (\ref{P_distributed}) to simplify the optimization, while keeping the second-order term to preserve the distributed network's structural information.
		Thus, (\ref{P_distributed}) can be approximated as $\mathbf{P} \approx \mathbf{O}^{H}\otimes \mathbf{O}^{H}$.
		
		In this case, we can reformulate the performance metric in \eqref{MSE_f2} for distributed sensing as follows
		\begin{equation}
			\label{distributed_problem}
			\begin{aligned}
				\mathcal{F}_{2}=
				\Big[& \alpha_{1} \left({s}^{e}\right)^{2} \operatorname{rvec}\{\mathbf{I}_{N}\otimes \mathbf{w}\mathbf{w}^{H}\}  \left(\mathbf{U}^{H} \otimes  \mathbf{U}^{H}\right)\\
				&+\operatorname{rvec}\{\mathbf{B}\} \Big]
				\mathbf{V} \Big\{ \mathbf{I}_{K^{2} N^{2}}- \mathbf{O}^{H}\otimes \mathbf{O}^{H} \Big\}^{-1}\operatorname{vec}\{ \mathbf{I}_{KN}\}.
			\end{aligned}
		\end{equation}
		
		Since the value range of $\mathcal{F}_1$ and $\mathcal{F}_2$ are different, it is desirable to normalize these two performance metrics in order to coordinate the sensing and communication capabilities, which can enhance the convergence of the ISAC system.
		Therefore, we define
		\begin{equation}
			\begin{aligned}
				\Psi_{p} = (\mathcal{F}_p - \mathcal{F}^{*}_p)/|\mathcal{F}^{*}_p|, p = 1,2,
			\end{aligned}
		\end{equation}
		where $\mathcal{F}^{*}_p, p =1,2$ is the corresponding performance limit of communication or sensing, which can be obtained by maximizing the communication SINR $\mathcal{F}_1$ or minimizing the sensing MSE $\mathcal{F}_2$ with power constraint, respectively.
		The corresponding optimization problems are convex, which can be easily solved by the CVX toolbox.
		
		In ISAC, it is expected to enhance the quality of communication and reduce sensing errors as much as possible with the limited resources to improve system performance \cite{zhang2022accelerating}.
		Therefore, we minimize the negative weighted sum of the normalized communication SINR $\Psi_{1}$ and the normalized sensing networked MSE $\Psi_{2}$ by optimizing the beamforming vectors $\mathbf{f}$ and $\mathbf{w}$ at the BS subject to the maximum transmit power constraint.
		The problem formulation can be expressed as
		\begin{equation}
			\label{opt_for_ISAC}
			\begin{aligned}
				&\mathop{\min}\limits_{\mathbf{f},\mathbf{w}} \quad -\beta_1 \Psi_{1}
				+\beta_2 \Psi_{2}\\
				&\text { s.t. } \quad \|\mathbf{w}\|^2+\|\mathbf{f}\|^2 \leq P,
			\end{aligned}
		\end{equation}
		where $P$ denotes the maximal transmit power of the BS, $\beta_p \geq 0$ is the priority of $\mathcal{F}_p$, which depends on the performance of the system designs and satisfies $\beta_1+\beta_2 = 1$. Note that the beamforming design above can effectively capture the impact of network topology on optimization performance due to the involvement of the adjacency matrix $\mathbf{C}$.
		
		In order to solve the problem in \eqref{opt_for_ISAC}, let us define $\mathbf{W} = \mathbf{w}\mathbf{w}^{H}$, $\mathbf{F} = \mathbf{f}\mathbf{f}^{H}$, $\mathbf{G}_{d} = |s^{d}|^2\mathbf{g}\mathbf{g}^{H}$, $\mathbf{G}_{e} = |s^{e}|^2\mathbf{g}\mathbf{g}^{H}$with $\operatorname{rank}\left(\mathbf{F}\right) = 1$ and $\operatorname{rank}\left(\mathbf{W}\right) = 1$. Note that such unit-rank
		constraints are non-convex.
		To eliminate this difficulty, we adopt the semi-definite relaxation (SDR) by omit the non-convex rank-1 constraint. Then, we integrate $\mathbf{W}$ and $\mathbf{F}$ into a matrix $\mathbf{Z} =
		\begin{bmatrix}
			\mathbf{W} & \mathbf{0}\\
			\mathbf{0} & \mathbf{F}
		\end{bmatrix}
		$. In this case, we have $\mathbf{W} =  \mathbf{I}_{w}^{H} \mathbf{Z} \mathbf{I}_{w}$, and
		$\mathbf{F}  = \mathbf{I}_{f}^{H} \mathbf{Z} \mathbf{I}_{f}$, where $\mathbf{I}_{w} = \begin{bmatrix}
			\mathbf{I}_{M} \\
			\mathbf{0}
		\end{bmatrix} \in \mathbb{C}^{2M \times M}$ and $\mathbf{I}_{f} = \begin{bmatrix}
			\mathbf{0} \\
			\mathbf{I}_{M}
		\end{bmatrix} \in \mathbb{C}^{2M \times M}$.
		Thus, problem \eqref{opt_for_ISAC} can be converted into an optimization problem with only one variable $\mathbf{Z}$ as
		\begin{equation}
			\label{optimal_for_matrix_1}
			\begin{aligned}
				&\mathop{\min}\limits_{\mathbf{Z}} \quad -\beta_1 \Psi_{1}\left( \mathbf{Z} \right)
				+\beta_2 \Psi_{2}\left( \mathbf{Z} \right) \\
				&\text{s.t.} \quad \operatorname{Tr}\left\{\mathbf{I}_{w}^{H}\mathbf{Z}\mathbf{I}_{w}\right\}+\operatorname{Tr}\left\{\mathbf{I}_{f}^{H}\mathbf{Z}\mathbf{I}_{f}\right\} -P  \leq 0.
			\end{aligned}
		\end{equation}
		
		Note that the objective function of \eqref{optimal_for_matrix_1} is the difference of convex (DC) functions, which lead to a NP-hard problem. To formulate address this issue, we introduce a penalty function for the constraint of \eqref{optimal_for_matrix_1} as
		\begin{equation}
			\label{penalty_function}
			\begin{aligned}
				p^{+}\left(\mathbf{Z}\right) = \operatorname{max}\left\{ \operatorname{Tr}\left\{\mathbf{I}_{w}^{H}\mathbf{Z}\mathbf{I}_{w}\right\}+\operatorname{Tr}\left\{\mathbf{I}_{f}^{H}\mathbf{Z}\mathbf{I}_{f}\right\}-P,0  \right\}.
			\end{aligned}
		\end{equation}
		By employing the penalty method \cite{pently}, \eqref{optimal_for_matrix_1} can be reformulated as the following unconstrained problem
		\begin{equation}
			\label{optimal_for_matrix_DC_penalty}
			\begin{aligned}
				\mathop{\min}\limits_{\mathbf{Z}} \varphi_{t}\left(\mathbf{Z}\right) &=  -\beta_1 \Psi_{1}\left( \mathbf{Z} \right)
				+\beta_2 \Psi_{2}\left( \mathbf{Z} \right)
				+ \delta_{t} p^{+}\left(\mathbf{Z}\right)\\
				&= g_{t}(\mathbf{Z}) - h_{t}(\mathbf{Z}).
			\end{aligned}
		\end{equation}
		where $g_{t}(\mathbf{Z}) =  \beta_2 \Psi_{2}\left( \mathbf{Z} \right) +  \delta_{t} p^{+}\left(\mathbf{Z}\right)$, and $h_{t}(\mathbf{Z}) = \beta_1 \Psi_{1}\left( \mathbf{Z} \right)$.
		Then, a general difference of convex algorithm (DCA) proposed in \cite{van2014advanced} can be used to obtain an effective suboptimal solution. In particular, the solution to the problem (\ref{optimal_for_matrix_DC_penalty}) is summarised in Algorithm \ref{alg:algorithm_DCA}.
		
		\begin{algorithm}[t]
			\caption{General DC Algorithm with Updated Penalty Parameter}  \label{alg:algorithm_DCA}
			\textbf{Initialize:} Take an initial matrix $\mathbf{Z}^{1} \in \mathbb{C}^{2M\times 2M}$; $\epsilon>0$; an initial penalty parameter $\delta_{1}>0$;
			the maximum number of iterations T and set $t=1$.
			
			\begin{algorithmic}[]
				\For{$t = 1:T$}
				\State  Compute $\mathbf{X}^{t} = \partial h_{t}\left( \mathbf{Z}^{t} \right)$.
				\State  Compute $ \mathbf{Z}_{t+1} = \operatorname{argmin}_{\mathbf{Z}} \left\{g_{t}(\mathbf{Z})-\left\langle \mathbf{Z}, \mathbf{X}^{t}\right\rangle \right\} $ by CVX toolbox.
				\State  if $\mathbf{Z}^{t+1} = \mathbf{Z}^{t}$ and $p\left( \mathbf{Z}^{t}\right)  \leq 0$
				\Statex	$\qquad \quad $ Return $\mathbf{Z}^{t}$.
				\State  Compute $r_{t}=\min \left\{\varphi_t\left(\mathbf{Z}^{t}\right), \varphi_t\left(\mathbf{Z}^{t+1}\right)\right\} $ and set
				\Statex	$\qquad \quad \delta_{t+1}\!=\! \begin{cases}\delta_{t}+\epsilon &\text{if } \delta_{t}\left\|\mathbf{Z}^{t+1}\!-\!\mathbf{Z}^{t}\right\|\!< \!1  \text{ and } r_{t}\!>\!0,\\
					\delta_{t} &\text {otherwise. }\end{cases}$
				\EndFor
				\State Return $\mathbf{Z}^{T}$.
			\end{algorithmic}
		\end{algorithm}

		Then, we can obtain the solution
		$\mathbf{Z}_\mathrm{opt} =
		\begin{bmatrix}
			\mathbf{W}_\mathrm{opt} & \mathbf{0}\\
			\mathbf{0} & \mathbf{F}_\mathrm{opt}
		\end{bmatrix}
		$. However, $\operatorname{rank}\left(\mathbf{W}_\mathrm{opt}\right) = 1$ and $\operatorname{rank}\left(\mathbf{F}_\mathrm{opt}\right) = 1$ cannot be guaranteed,
		which means that we cannot directly obtain the optimal vectors $\mathbf{w}_\mathrm{opt}$ and $\mathbf{f}_\mathrm{opt}$ from $\mathbf{W}_\mathrm{opt}$ and  $\mathbf{F}_\mathrm{opt}$.
		Instead, we determine an approximation $\mathbf{w}_\mathrm{opt}$ and $\mathbf{f}_\mathrm{opt}$ using the Gaussian randomization method:
		
		1) For $g \in\{0,1, \ldots, G-1\}=\mathcal{G}$, generate $G$ random vectors $\boldsymbol{\nu}_{g} \sim \mathcal{C} \mathcal{N}\left(0, \mathbf{W}_{\text {opt }}\right)$, $\boldsymbol{\xi}_{g} \sim \mathcal{C} \mathcal{N}\left(0, \mathbf{F}_{\text {opt }}\right)$ .
		
		2) For $g \in \mathcal{G}$, set $\boldsymbol{\nu}_{g}, \boldsymbol{\xi}_{g} \geq \mathbf{0}$.
		
		3) $\hat{\mathbf{w}}_{\mathrm{opt}} , \hat{\mathbf{f}}_{\mathrm{opt}}  =\arg \min _{\boldsymbol{\nu}_{g},\boldsymbol{\xi}_{g} \forall g \in \mathcal{G}} -\beta_1 \Psi_{1}+\beta_2 \Psi_{2}$.
		
		By applying the Gaussian randomization, we refer to $\hat{\mathbf{w}}_{\mathrm{opt}}$ and $ \hat{\mathbf{f}}_{\mathrm{opt}} $ as the optimized solutions.
		
		\section{Numerical Results}\label{section_5}
		In this section, we provide numerical results to validate the performance of our proposed two-step distributed sensing algorithm and resource allocation algorithm for the ISAC system and then discuss the impacts of different system parameters on the system.
		All simulations are executed using MATLAB 2020b on a dedicated computing server equipped with an Intel(R) Xeon(R) E5-2680 v4 processor and 64 GB of memory.

		Unless otherwise stated, to represent the sparse structure of the ROI, there are only a few consecutive non-empty pixels in the ROI, and all other pixels are empty.
		The simulation scenario is set in a room with a size of $4\text{m}\times4\text{m} \times4\text{m}$, and the surrounding environment information is depicted through $4\times4\times4$ small sub-cubes, which serve to represent both the spatial distribution and scattering coefficients.
		The scattering coefficients are set as $x_k \geq 0, \forall k \in \{ 1, \dots, K\}$, where element 0 means the corresponding position in the ROI is a empty pixel.
		The number of BS antennas is 16, and the maximal transmit power $P$ is 10 W.
		The transmitted signal frequency is set to 28 GHz.
		We consider a network with 20 sensing users, each user connects to three users on average and randomly distributed near the scatter.
		For simplicity, we set the same Gaussian noise distribution for all devices in the environment, which means that the variance of the noise is the same for different devices, i.e., $\sigma_{\mathrm{o,n}}^2 = \sigma^2, \forall n $.
		We use $\operatorname{SNR} = 10 \log_{10}(P/\sigma^2)$ to denote the transmit signal-to-noise ratio (SNR) (in dB) at the sensing users.
		The coefficient of the linear combination $\mathbf{C}$ in \eqref{DTLS_loc_functon1} is constructed by the Metropolis rule in this paper, which has been widely adopted in other distributed estimation algorithms \cite{liu2012diffusion,shao_coop, shao2019complementary}, i.e.,
		\begin{equation}
			c_{l k}= \begin{cases}\frac{1}{\max \left(n_{k}, n_{l}\right)}, & l \in N_{k} \backslash\{k\}, \\ 1-\displaystyle\sum_{l \in N_{k} \backslash\{k\}} c_{l k}, & l=k, \\ 0, & l \notin N_{k},\end{cases}
		\end{equation}
		where $n_k$ and $n_l$ are the degrees (numbers of links) of users $n$ and $l$, respectively. $N_{k} \backslash\{k\}$ is the index set of the neighbors of the $k$-th user except itself.
		The network mean-square-deviation (MSD) is employed in the experimental simulations \cite{liu2012diffusion}, indicating the difference between the estimate $\mathbf{x}_{k,i}$ of each user $k$ and optimal ROI $\mathbf{x}^{0}$ at instant $i$, which is define as
		\begin{equation}
			\operatorname{MSD}(i) = 10\operatorname{log}\left(\frac{1}{N}
			\sum_{k=1}^N
			\operatorname{E} \left[ \left\| \mathbf{x}^{0} - \mathbf{x}_{k,i}\right\|_{2}^{2} \right] \right).
		\end{equation}
		
		\begin{figure}[t!]
			\setlength{\abovecaptionskip}{-0.cm}
			\setlength{\belowcaptionskip}{0.cm}
			\centering
			\includegraphics [width=0.48\textwidth] {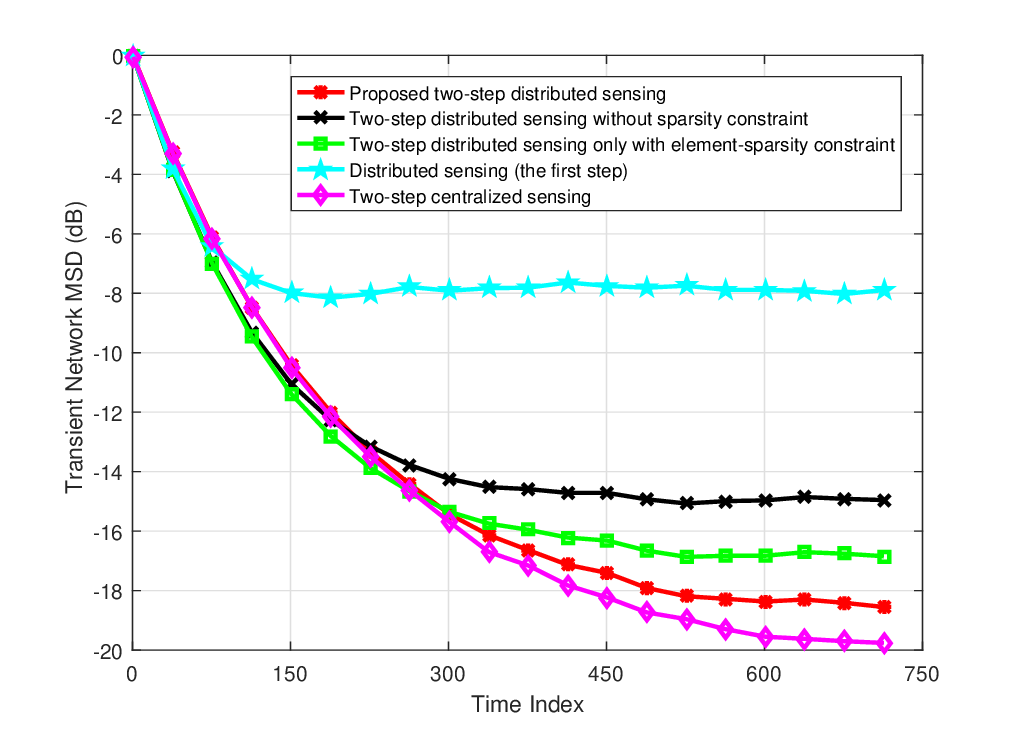}
			\caption{Transient network MSDs of proposed two-step distributed sensing, two-step distributed sensing without sparsity constraint, two-step distributed sensing with element-sparsity, distributed sensing (the first step), and two-step centralized sensing.}
			\label{MSD_Transient}
		\end{figure}

		First, the convergence of the proposed two-step distributed sensing, the two-step distributed sensing without sparsity constraint (i.e., without the $l_{1}$-norm and the $l_{21}$-norm penalties in \eqref{DTLS_loc_functon1}), the two-step distributed sensing only with element-sparsity (i.e., without the $l_{1}$-norm penalty in \eqref{DTLS_loc_functon1}), the two-step centralized sensing, and the proposed distributed sensing only with step 1 of Algorithm \ref{alg:algorithm_SDTLS_two_step}, are given in Fig. \ref{MSD_Transient}.
		In the performance comparison of distributed sensing strategies with and without sparsity constraint, we can observe that the MSD of the black curve is higher than that of the green and red lines.
		This indicates that sparsity penalty enables the distributed sensing strategy with a lower sensing MSD.
		Moreover, the transient network MSD of the proposed two-step distributed sensing algorithm is lower than that of the two-step distributed sensing with element sparsity.
		This is because the proposed two-step distributed sensing explores not only element sparsity but also block-wise sparsity.
		Importantly, the proposed two-step distributed sensing performs better than the first-step distributed sensing algorithm by cancelling interference from received signals, leading to an estimate of the ROI that is closer to the actual one.
		Notably, while the proposed two-step distributed sensing suffers performance losses compared with the two-step centralized strategy due to incomplete data collection in the distributed algorithm, it generally enjoys faster convergence.

		\begin{figure}[t]
			\centering
			\vspace{-20pt}
			\subfigbottomskip=1pt
			\subfigcapskip=-10pt
			\subfigure[The original ROI.]{ \label{ROI_subfigure_1}
				\includegraphics[width=0.45\linewidth]{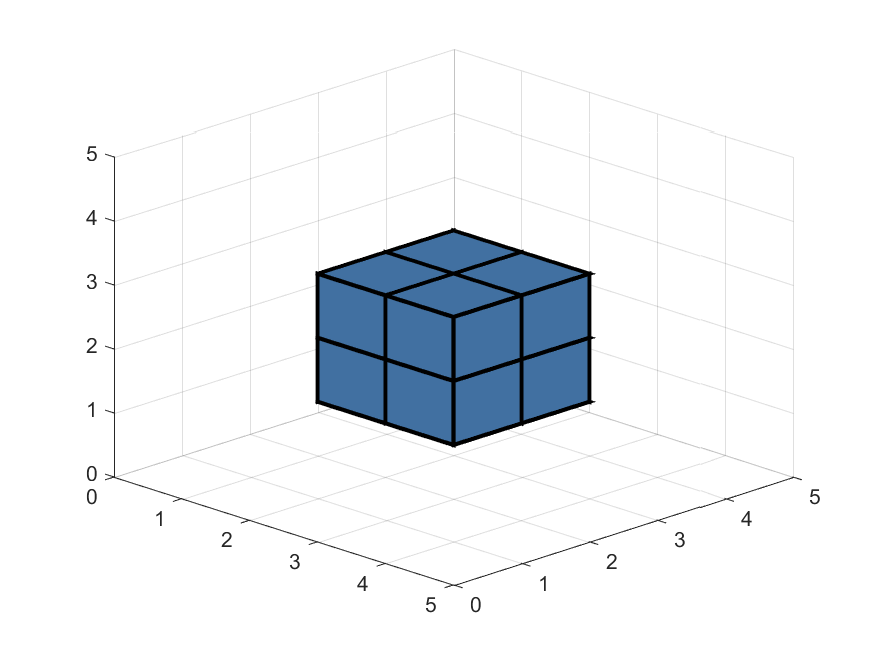}}
			\subfigure[The sensing result with $\operatorname{SNR} = 3\operatorname{dB}$]{ \label{ROI_subfigure_2}
				\includegraphics[width=0.45\linewidth]{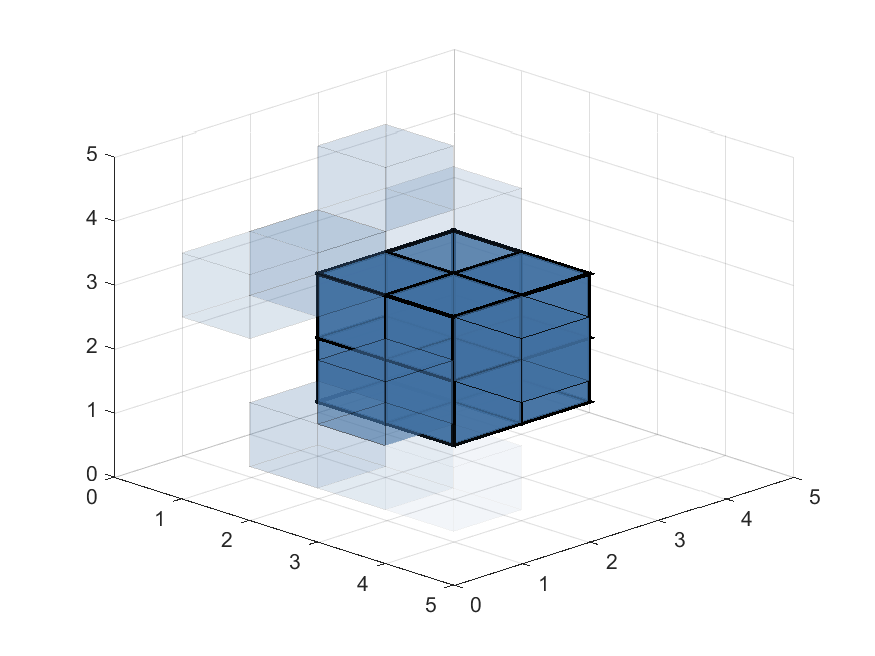}}
			\\
			\subfigure[The sensing result with $\operatorname{SNR} = 5\operatorname{dB}$]{ \label{ROI_subfigure_3}
				\includegraphics[width=0.45\linewidth]{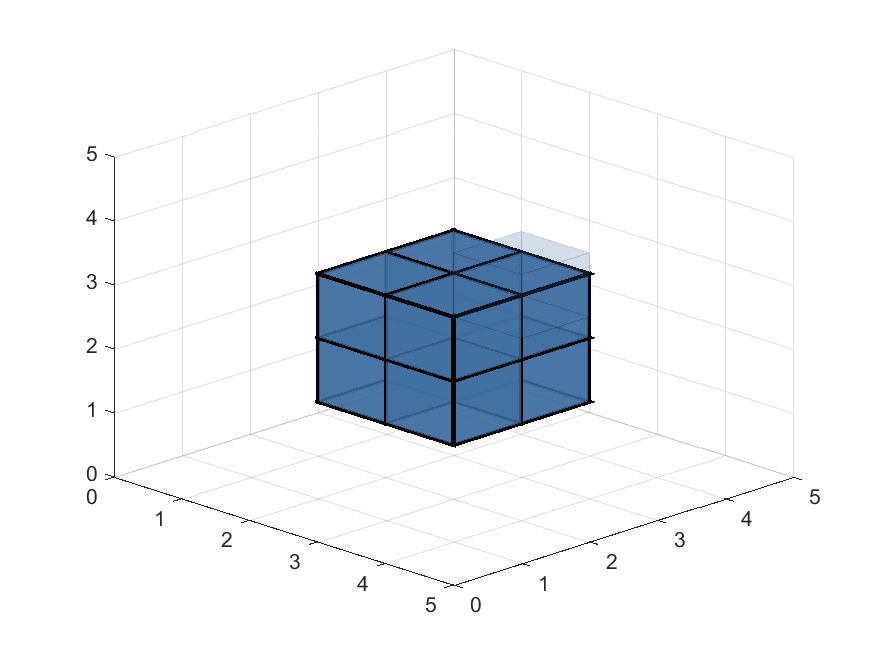}}
			\subfigure[The sensing result with $\operatorname{SNR} = 10\operatorname{dB}$]{ \label{ROI_subfigure_4}
				\includegraphics[width=0.45\linewidth]{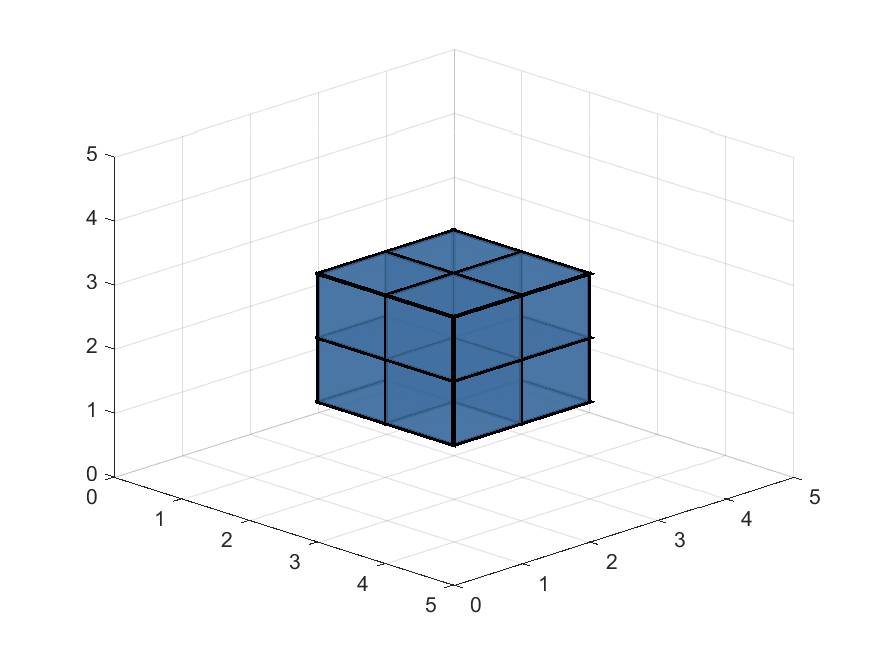}}
			\caption{The original ROI and the sensing results under different SNR conditions.
				The transparency of the small sub-cube represents the estimated scattering coefficient, the more transparent the small sub-cube is, the smaller the corresponding scattering coefficient is.}
			\label{ROI_sensing_result}
		\end{figure}

	When our proposed iterative algorithm converges, the intuitive sensing results are shown in Figs. \ref{ROI_subfigure_2}, \ref{ROI_subfigure_3}, and \ref{ROI_subfigure_4} with the SNR being $3  \operatorname{dB}$, $5 \operatorname{dB}$, and $10 \operatorname{dB}$, respectively.
	We denote the coefficients by transparency, the lower the transparency of the small sub-cube, the larger the scattering coefficient of the point.
	It can be seen that when $\operatorname{SNR} = 3\operatorname{dB}$ in Fig. \ref{ROI_subfigure_2}, there are many small discrete squares with shadows. This indicates that sensing in a 3 dB noise environment causes many errors, and it is difficult to distinguish the shape of the ROI from the ambient noise.
	Although when $\operatorname{SNR} = 5\operatorname{dB}$ in Fig. \ref{ROI_subfigure_3}, the shape of the ROI can be clearly distinguished, the sensing result retains some environmental noise.
	As the SNR further increases, the number of incorrectly identified pixels gradually decreases, and when $\operatorname{SNR} = 10\operatorname{dB}$ in Fig. \ref{ROI_subfigure_4}, the sensing result can clearly distinguish the shape of the target.
	This is because the intensity of the environmental noise affects the accuracy of communication signal cancellation, which in turn has an impact on the effectiveness of the target sensing results.
	
	\begin{figure}[t!]
		\setlength{\abovecaptionskip}{-0.cm}
		\setlength{\belowcaptionskip}{0.cm}
		\centering
		\includegraphics [width=0.48\textwidth] {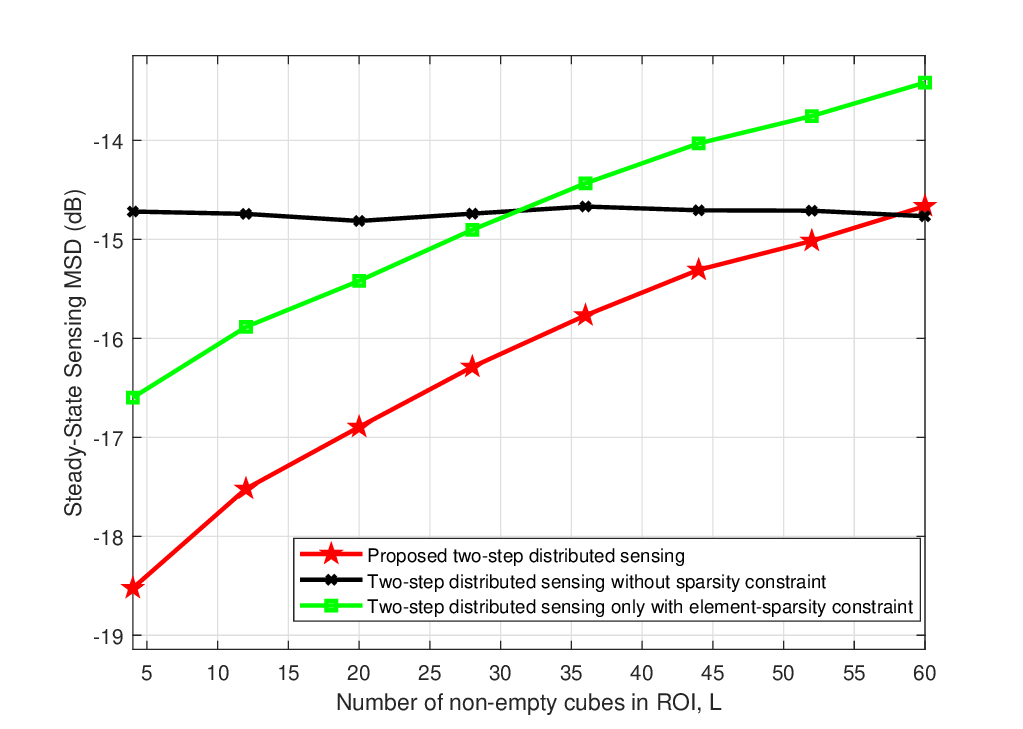}
		\caption{Steady-state sensing MSDs against different sparsities of ROI with $K = 64, N = 20$, and $ \beta_{1}= 0.1$. }
		\label{result_sparse}
	\end{figure}

		\begin{figure}[t!]
			\setlength{\abovecaptionskip}{-0.cm}
			\setlength{\belowcaptionskip}{0.cm}
			\centering
			\includegraphics [width=0.48\textwidth] {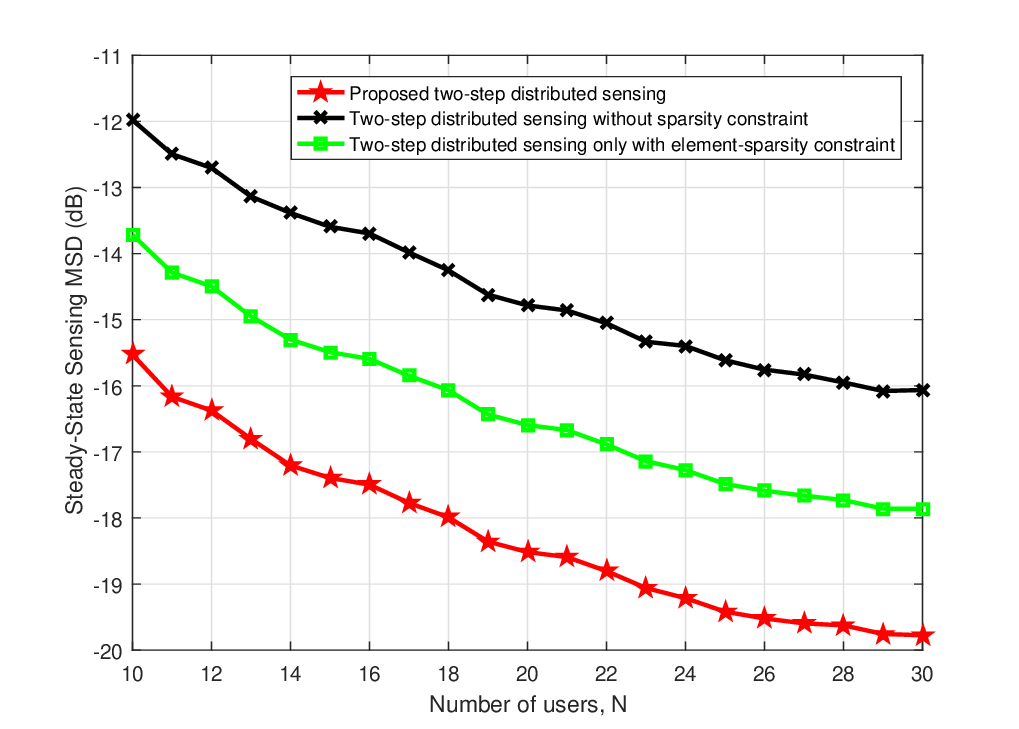}
			\caption{Steady-state sensing MSDs versus the number of users with $K = 64, L = 8$, and $ \beta_{1}= 0.1$.}
			\label{result_Node}
		\end{figure}

		Then, we investigate the influence of sparsity on the sensing performance by measuring the sensing steady-state MSDs.
		In our simulations, we obtain the steady-state MSD by averaging the results in the last 150 samples after 600 iterations of 50 simulations.
		The superiority of our proposed two-step distributed sensing strategy in steady-state MSD performance is more pronounced when the ROI is sparse, as shown in Fig. \ref{result_sparse}.
		As the number of non-zero components (i.e., $L = \|x\|_0$) grows larger, such superiority gradually weakens and eventually vanishes when the ROI is entirely non-sparse.
		In addition, when $L > 30$, the steady-state MSD of the distributed sensing strategy with element sparsity is higher compared to the case without sparsity constraint.
		This is because the value of the given regularization parameter exceeds the range of the regularization parameter for the distributed sensing strategy with element-sparsity in this case, which is also illustrated in \cite{DTLS-Li}.
		For the same reason, in the case of $L>56$, we can see that the steady-state MSD of our proposed two-step distributed sensing strategy is slightly higher than the case without sparsity constraint.
		
		\begin{figure}[t]
			\setlength{\abovecaptionskip}{-0.cm}
			\setlength{\belowcaptionskip}{0.cm}
			\centering
			\includegraphics [width=0.48\textwidth] {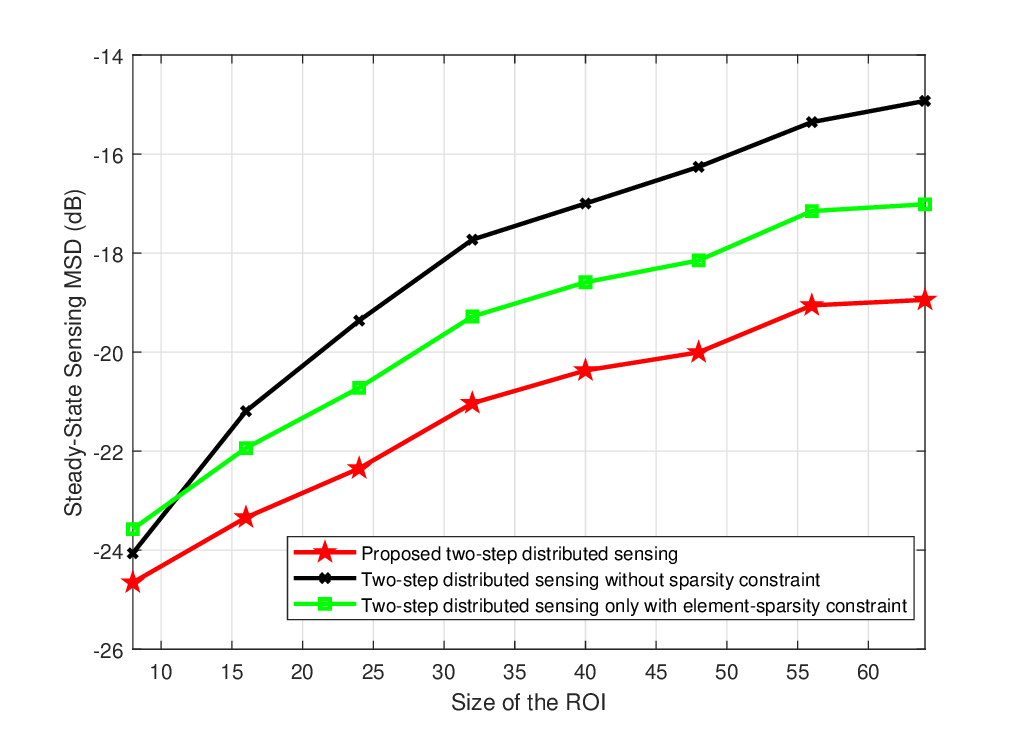}
			\caption{Steady-state sensing MSDs versus the size of ROI with $L/K = 1/8, N = 20 $, and $ \beta_{1}= 0.1$.}
			\label{result_ROI}
		\end{figure}
		
		In Fig. \ref{result_Node}, we discuss the impact of the number of users on the sensing accuracy, while keeping the number of non-zero elements and the size of the ROI constant.
		We can see that the ability of distributed sensing gradually increases as the number of users increases, indicating that a larger number of users is beneficial for distributed sensing results.
		This is due to the fact that an increase in the number of users enhances the collaborative capabilities and measurement diversity, which are employed together to improve the imaging capabilities.

		In Fig. \ref{result_ROI}, we analyze the impact of the size of the ROI on the performance of sensing results, where the ratio of the number of non-zero elements of the ROI to the size of the ROI is fixed at $\frac{L}{K} = \frac{1}{8}$.
		It is illustrated that the sensing capability of the system gradually decreases as the size of the ROI increases. This is because the larger the target, the more scattering coefficients, resulting in more serious double fading of path loss, thus reducing the received SNR and target recovery accuracy.
		\begin{figure}[t!]
			\setlength{\abovecaptionskip}{-0.cm}
			\setlength{\belowcaptionskip}{0.cm}
			\centering
			\includegraphics [width=0.48\textwidth] {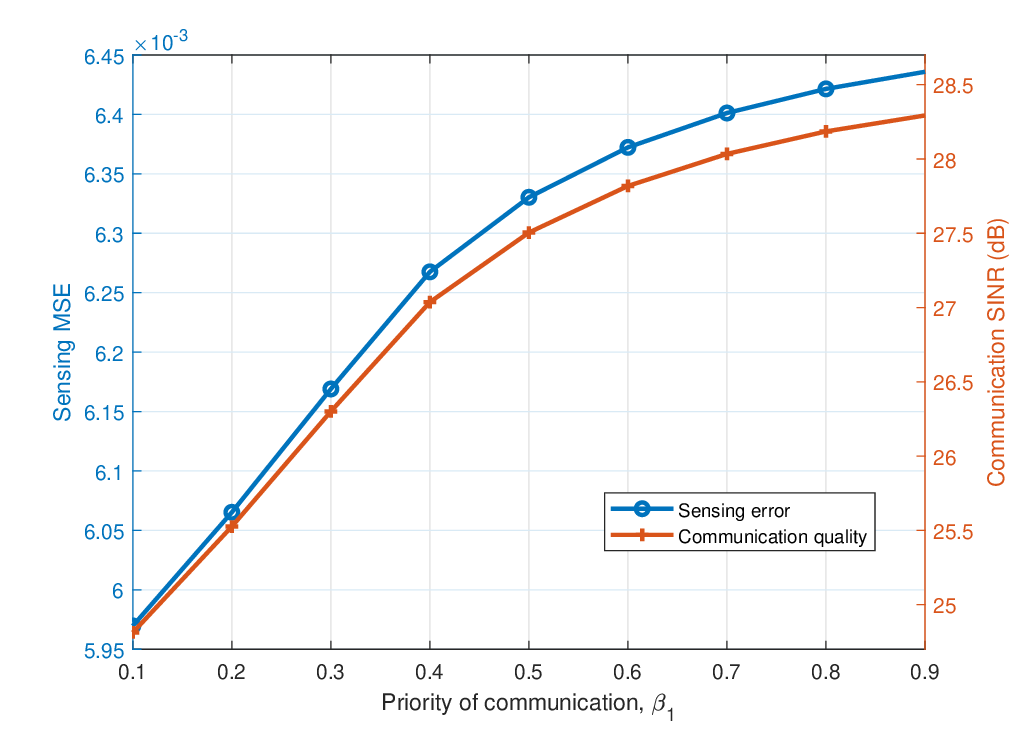}
			\caption{The system performance versus the priorities of sensing and communication with $K = 64, L = 8$, and $ N= 20$. }
			\label{result_opti}
		\end{figure}
		
		\begin{figure}[t!]
			\setlength{\abovecaptionskip}{-0.cm}
			\setlength{\belowcaptionskip}{0.cm}
			\centering
			\includegraphics [width=0.48\textwidth] {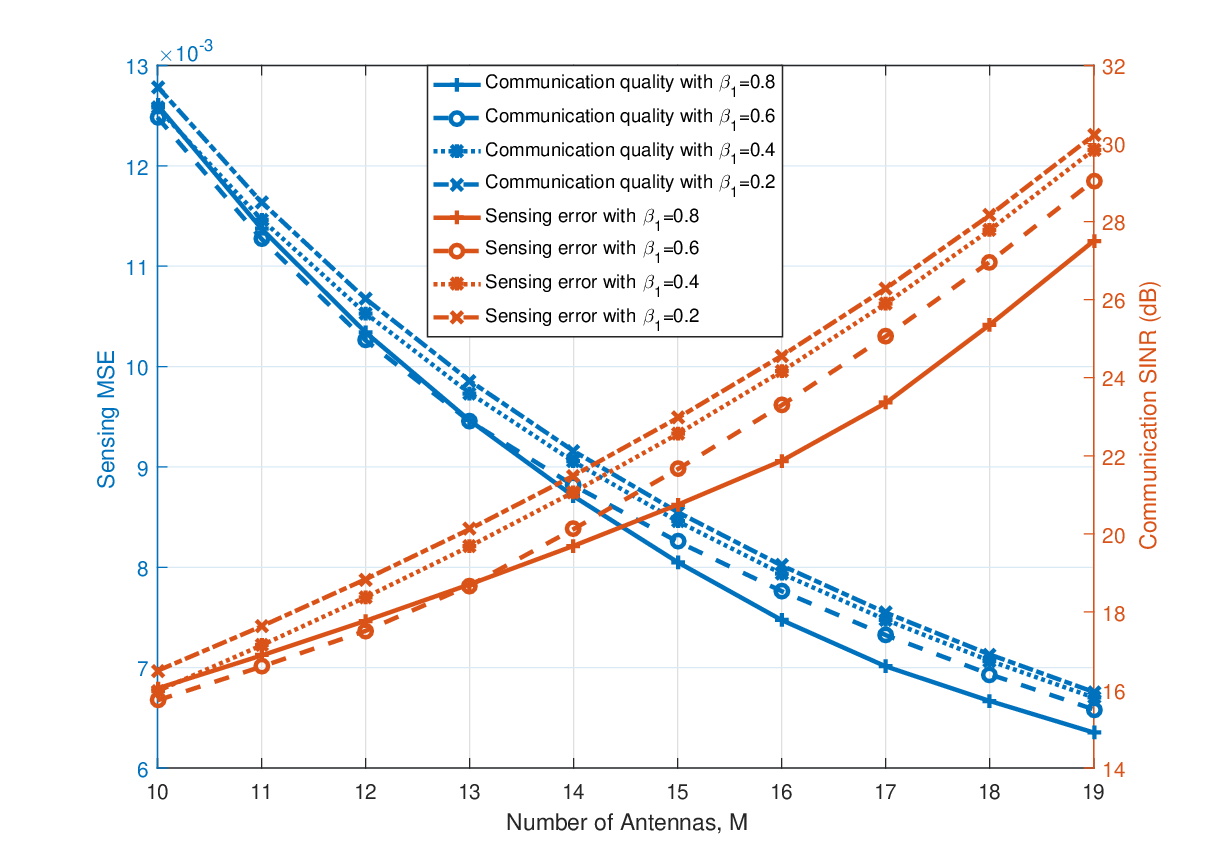}
			\caption{The system performance gains versus the number of antennas for Algorithm \ref{alg:algorithm_SDTLS} with different coefficients with $K = 64, L = 8$, and $N= 20$.}
			\label{gain_with_M}
		\end{figure}

		Next, we explore the impact of the priorities of sensing and communication on system performance.
		We obtain different results for the system by varying the priority $\beta_1$ of the communication function.
		Fig. \ref{result_opti} depicts the performance of the two functions with various priorities.
		It can be observed that as the priority of communication increases, the communication quality of the system gradually increases, while the capability of sensing gradually decreases.
		Therefore, it is meaningful to choose an appropriate set of priorities to balance the performance of sensing and communication according to the actual situation.

		Finally, Fig. \ref{gain_with_M} shows the influence of the number of antennas $M$ at BS on ISAC performance with sensing priority $\beta_{1}$ and communication priority $\beta_{2}$.
		The more antennas the BS has, the lower the sensing error of the algorithm and the higher the quality of the communication, which means better system performance.
		This is because more antennas provide a higher spatial multiplexing gain, which can be exploited to effectively improve ISAC system performance.

		\section{Conclusion} \label{section_6}

		We designed a robust networked ISAC system and proposed a two-step distributed sensing algorithm based on multi-user collaboration.
		We exploited the block-sparsity of the imaging ROI and interference cancellation with TLS to improve the sensing accuracy and discuss the effect of system parameters on ISAC performance.
		The beamforming was designed by minimizing the sensing error and maximizing communication quality with limited power constraint.
		Importantly, depending on the application preference of the wireless network in realistic scenarios, it is feasible to obtain the desired sensing and communication performance by altering the priority of the system.
		In addition, we found that multi-user collaboration can provide benefits to the ISAC system.
		Finally, numerical simulations verify the convergence and effectiveness of our proposed scheme.
		
		\begin{appendices}
			
			\section{Derivation of $\mathbf{q}$ and $\mathbf{P}$}\label{Appendix_A}
			In this appendix, we derive the expression of the vector $\mathbf{q}$ and the matrix $\mathbf{P}$ in \eqref{MSE_f2}.
			We replace ${\bf{u}}_{n,i}^{H}$ with ${s}_{i}^{\mathrm{e}}\mathbf{w}^H{\bf{G}}{\operatorname{diag}}({{\bf{h}}_n})$ in (\ref{U_li}) to rewrite ($\ref{R_l}$) as
			\begin{equation}
				\begin{aligned}
					\mathbf{R}_{n}=E\left[\mathbf{u}_{n, i} \mathbf{u}_{n, i}^{H}\right]
					=(s^{e})^{2}\operatorname{diag}(\mathbf{h}_{n})\mathbf{G}^{H}\mathbf{w}\mathbf{w}^{H}\mathbf{G}\operatorname{diag}(\mathbf{h}_{n}).
				\end{aligned}
			\end{equation}
			Based on this, we can then rewrite $\operatorname{diag}\{\mathbf{R}_{1},\dots,\mathbf{R}_{N}\}$ as
			\begin{equation}
				\begin{aligned}
					&\operatorname{diag}\{\mathbf{R}_{1},\dots,\mathbf{R}_{N}\} \\=&
					\left({s}^{e}\right)^{2}\operatorname{diag}\{ \operatorname{diag}\left(\mathbf{h}_{1}\right)  ,\dots, \operatorname{diag}\left(\mathbf{h}_{N}\right) \} \left( \mathbf{I}_{N} \otimes \mathbf{G}^{H}\right)\\ 
					& \! \times \!\left(\mathbf{I}_{N} \!\otimes\!  \mathbf{w}\mathbf{w}^{H}\right)\left( \mathbf{I}_{N} \!\otimes\! \mathbf{G}\right)\operatorname{diag}\{ \operatorname{diag}\left(\mathbf{h}_{1}\right) ,\dots, \operatorname{diag}\left(\mathbf{h}_{N}\right) \}\\
					=& 	\left({s}^{e}\right)^{2} \mathbf{U}  \left(\mathbf{I}_{N} \otimes  \mathbf{w}\mathbf{w}^{H}\right)  \mathbf{U}^{H},
				\end{aligned}
			\end{equation}
			where $\mathbf{U} = \operatorname{diag}\{ \operatorname{diag}\left(\mathbf{h}_{1}\right)  ,\dots, \operatorname{diag}\left(\mathbf{h}_{N}\right) \}\left( \mathbf{I}_{N} \otimes \mathbf{G}^{H}\right)$. Then, $\mathbf{Q}$ in (\ref{Q}) can be rewritten as
			\begin{equation}
				\begin{aligned}
					\label{q}
					\mathbf{Q} \!=\!& \operatorname{diag}\{\!E\!\left[\mathbf{m}_{1}\!\left(\mathbf{x}^{0}\right)\! \mathbf{m}_{1}^{H}\!\left(\mathbf{x}^{0}\right)\right],\dots, \!E\!\left[\mathbf{m}_{N}\!\left(\mathbf{x}^{0}\right) \mathbf{m}_{N}^{H}\!\left(\mathbf{x}^{0}\right)\right] \}\\
					=& \frac{4 \boldsymbol{\sigma}_{\text { in }}^{2}}{\left\|\mathbf{x}^{0}\right\|^{2}+1}\operatorname{diag}\{\mathbf{R}_{1},\dots,\mathbf{R}_{N}\}\\
					&\qquad +\mathbf{I}_{N} \otimes \left( \boldsymbol{\sigma}_{\text{ in }}^{2} \mathbf{I}_{K}
					-\frac{3 \boldsymbol{\sigma}_{\text { in }}^{2} \mathbf{x}^{0}\left(\mathbf{x}^{0}\right)^{H}}{\left\|\mathbf{x}^{0}\right\|^{2}+1}\right)\\
					=& \alpha_{1} \left({s}^{e}\right)^{2} \mathbf{U}  \left(\mathbf{I}_{N} \otimes  \mathbf{w}\mathbf{w}^{H}\right)  \mathbf{U}^{H}
					+\mathbf{B},
				\end{aligned}
			\end{equation}
			where $\alpha_{1} = \frac{4 \boldsymbol{\sigma}_{\text { in }}^{2}}{\left\|\mathbf{x}^{0}\right\|^{2}+1}$, $\mathbf{B} = \mathbf{I}_{N} \otimes \left( \boldsymbol{\sigma}_{\text{ in }}^{2} \mathbf{I}_{K}
			-\frac{3 \boldsymbol{\sigma}_{\text { in }}^{2} \mathbf{x}^{0}\left(\mathbf{x}^{0}\right)^{H}}{\left\|\mathbf{x}^{0}\right\|^{2}+1}\right)$.
			Based on \eqref{q}, the vector $\mathbf{q}^{H}$ can be rewritten as
			\begin{equation}
				\begin{aligned}
					\mathbf{q}^{H} =&\{\operatorname{vec}\{\mathbf{Q}\}\}^{H}\\
					=& \{\alpha_{1} \left({s}^{e}\right)^{2}\left(\mathbf{U}\otimes \mathbf{U}\right)
					\operatorname{vec}\{\mathbf{I}_{N}\otimes \mathbf{w}\mathbf{w}^{H}\}
					+\operatorname{vec}\{\mathbf{B}\}\}^{H}\\
					=&\alpha_{1} \left({s}^{e}\right)^{2} \operatorname{rvec}\{\mathbf{I}_{N}\otimes \mathbf{w}\mathbf{w}^{H}\}  \left(\mathbf{U}^{H} \otimes  \mathbf{U}^{H}\right)
					+\operatorname{rvec}\{\mathbf{B}\} .
				\end{aligned}
			\end{equation}
			Finally, $\mathbf{P}$ in \eqref{p0} can be rewritten as
			\begin{equation}
				\label{P}
				\begin{aligned}
					\mathbf{P} &= \left[\left(\mathbf{I}_{K N}-\mathbf{D} \mathbf{H}\right) \mathbf{O}^{H} \right] \otimes \left[\left(\mathbf{I}_{K N}-\mathbf{D} \mathbf{H}\right) \mathbf{O}^{H} \right]\\
					&= \mathbf{O}^{H}\otimes \mathbf{O}^{H}-\left(\mathbf{H} \otimes \mathbf{I}_{K N}\right) \left[\left(\mathbf{D} \mathbf{O}^{H} \right) \otimes \mathbf{O}^{H}\right]\\
					&\;\;-\left( \mathbf{I}_{K N} \otimes \mathbf{H} \right)\left[\mathbf{O}^{H}\otimes \left(\mathbf{D}\mathbf{O}^{H} \right) \right]\\
					&\;\;+  \left( \mathbf{H} \otimes \mathbf{H} \right)\left( \mathbf{D} \otimes \mathbf{D}\right)\left( \mathbf{O}^{H} \otimes \mathbf{O}^{H}  \right)\\
					&=\mathbf{O}^{H}\otimes \mathbf{O}^{H}\\
					&\;\;-\!\alpha_{2}\! \left(\!{s}^{e}\!\right)^{2}\! \left\{\! \left[\mathbf{U}\!\left(\mathbf{I}_{N}\!\otimes\!  \mathbf{w}\mathbf{w}^{H}\right)\!\mathbf{U}^{H}  \right]\! \otimes \!\mathbf{I}_{KN}\! \right\}\! \left[\!\left( \mathbf{D}\!\mathbf{O}^{H}  \right)\!\otimes\! \mathbf{O}^{H}  \right]\!\\
					&\;\;-\! \alpha_{2}\! \left(\!{s}^{e}\!\right)^{2}\! \left\{\! \mathbf{I}_{KN}\! \otimes\! \left[\mathbf{U}\!\left(\mathbf{I}_{N}\!\otimes\!  \mathbf{w}\mathbf{w}^{H} \right)\!\mathbf{U}^{H}  \right]\right\}\! \left[ \mathbf{O}^{H} \!\otimes  \!\left( \mathbf{D}\mathbf{O}^{H}  \right)  \right] \\
					&\;\;+\!\alpha_{2}^{2}\! \left(\!{s}^{e}\!\right)^{4}\! \left\{\!\left[\mathbf{U}\!\left(\mathbf{I}_{N}\!\otimes\!  \mathbf{w}\mathbf{w}^{H} \right)\!\mathbf{U}^{H}  \right] \!\otimes\! \left[\mathbf{U}\!\left(\mathbf{I}_{N}\!\otimes \! \mathbf{w}\mathbf{w}^{H} \right)\!\mathbf{U}^{H}  \right]\! \right\}\\ 
					&\quad \times \left(\mathbf{D}\otimes \mathbf{D} \right) \left(\mathbf{O}^{H}\otimes \mathbf{O}^{H} \right),
				\end{aligned}
			\end{equation}
			where $\alpha_{2} =\frac{2}{\left\|\mathbf{x}^{0}\right\|_{2}^{2}+1} $ is a constant.

			\footnotesize
			\bibliographystyle{IEEEtran}
			\bibliography{IEEEabrv,refs}
		\end{appendices}
	\begin{IEEEbiography}[{\includegraphics[width=1in,height=1.25in,clip,keepaspectratio]{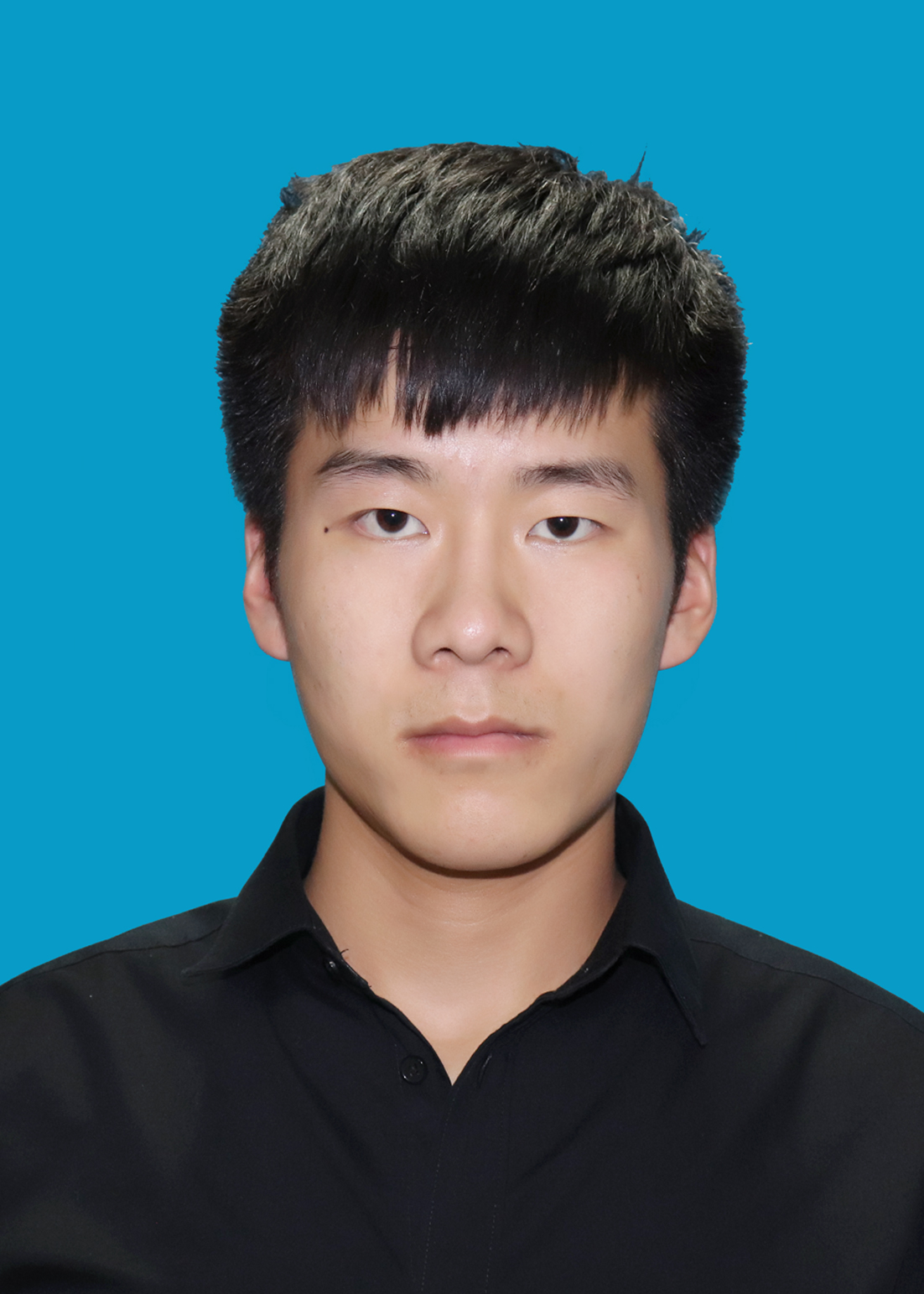}}]{Jiapeng Li}
		received the B.S. degree from Beijing University of Civil Engineering and Architecture, China, in 2020 and the M.S. degree from Southwest University, China, in 2023.
		He is currently pursuing the Ph.D. degree in mathematics at Southern University of Science and Technology.
		His research interests include adaptive signal processing,  integrated communication and sensing, and near-field communication.
	\end{IEEEbiography}
	\begin{IEEEbiography}[{\includegraphics[width=1in,height=1.25in,clip,keepaspectratio]{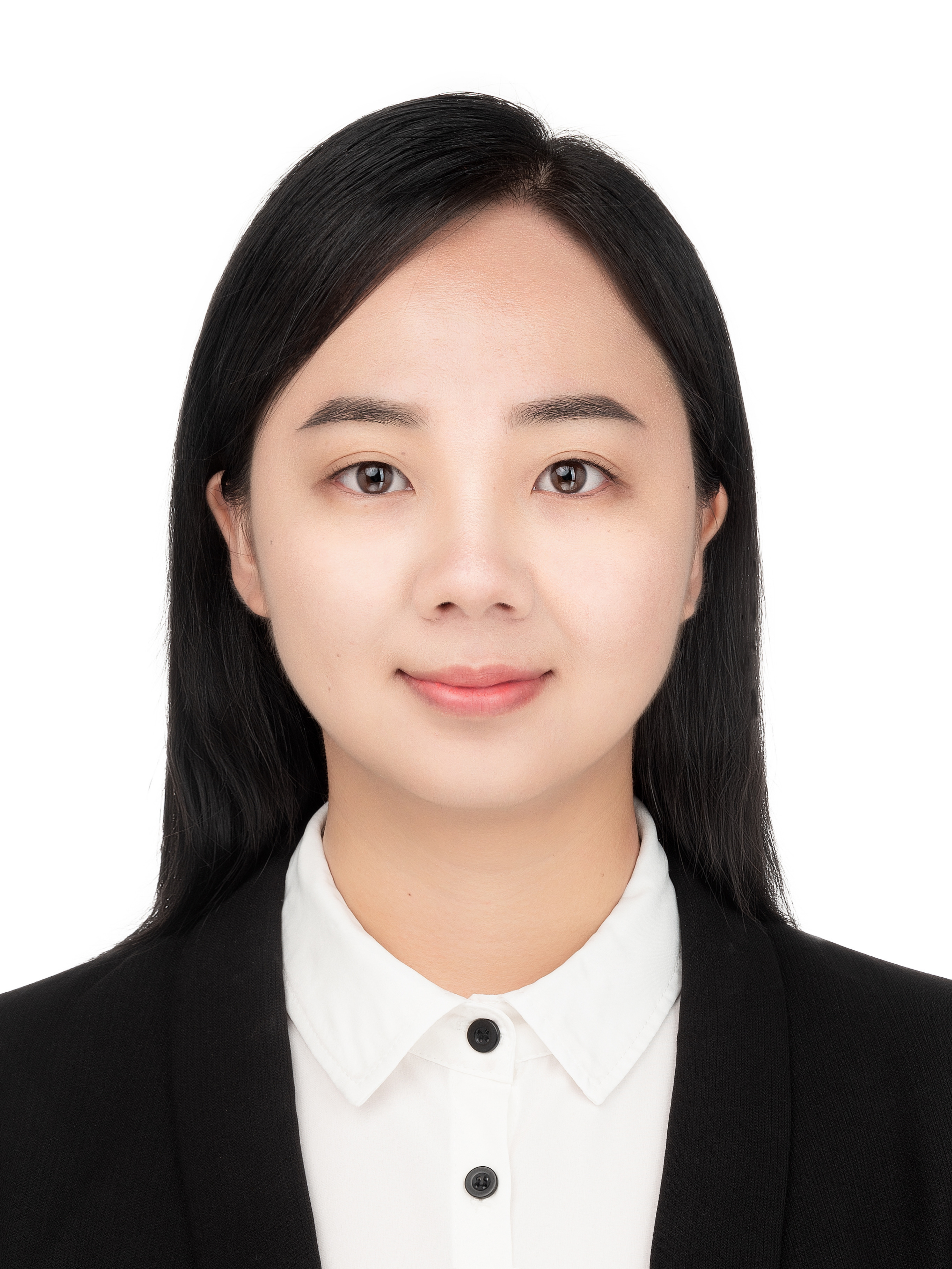}}]
		{Xiaodan Shao}(M'22) received the Ph.D. degree in information and communication engineering from Zhejiang University in 2022. She is currently a Humboldt Postdoctoral Research Fellow with the Institute for Digital Communications, Friedrich-Alexander-University Erlangen-Nuremberg (FAU), Erlangen, Germany. From June 2017 to June 2018, she was with the China Resources Microelectronics Ltd. as a Member of Technical Staff. From 2021 to 2022, she was a Visiting Research Scholar with the Department of Electrical and Computer Engineering, National University of Singapore. Her current research interests include 6DMA (six-dimensional movable antenna), massive access, and statistical signal processing. She was the recipient of the Best Ph.D. Thesis Award of the China Institute of Communications
		in 2023, the Best Ph.D. Thesis Award of Zhejiang Province in 2022, and the IEEE International Conference on Wireless Communications and Signal Processing (WCSP) Best Paper Award in 2020.
	\end{IEEEbiography}
	\begin{IEEEbiography}[{\includegraphics[width=1in,height=1.25in,clip,keepaspectratio]{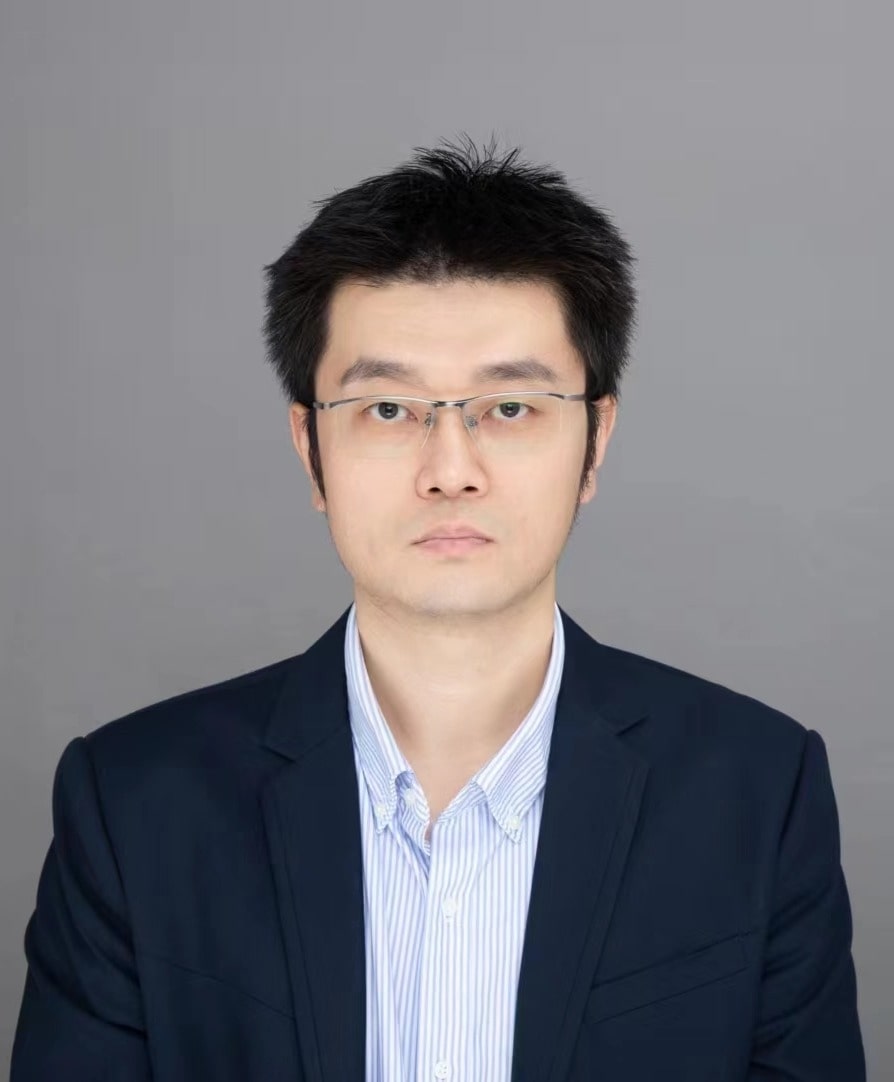}}]{Feng Chen}
		(Member, IEEE) received the Ph.D. degree from the University of Electronic Science and Technology of China in 2013. From 2008 to 2010, he visited in Prof. Emery Browns research group with Massachusetts Institute of Technology, and also worked with Harvard University, Massachusetts General Hospital. Currently, he is the vice dean and professor of the College of Artificial Intelligence,
		Southwest University, Chongqing, China. He has published over 50 articles in various journals and conference proceedings. His research interests include signal estimation, image denoising, signal filtering, and point cloud registration.
	\end{IEEEbiography}	
	\begin{IEEEbiography}[{\includegraphics[width=1in,height=1.25in,clip,keepaspectratio]{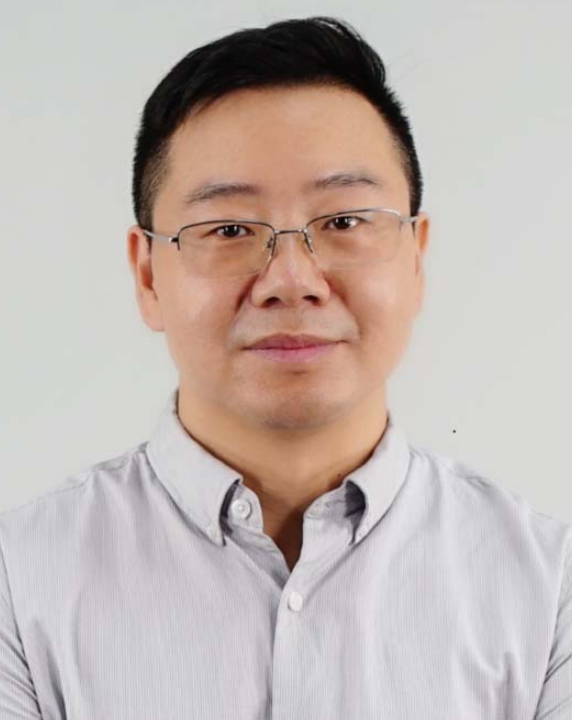}}]{Shaohua Wan}
		(Senior Member, IEEE) received the Ph.D. degree from the School of Computer, Wuhan University, in 2010. He is currently a Full Professor with the Shenzhen Institute for Advanced Study, University of Electronic Science and Technology of China. From 2016 to 2017, he was a Visiting Professor at the Department of Electrical and Computer Engineering, Technical University of Munich, Germany. He is the author of over 150 peer-reviewed research papers and books, including over 50 IEEE/ACM TRANSACTIONS papers, such as IEEE Transactions on Mobile Computing, IEEE Transactions on Industrial Informatics, IEEE Transactions on Intelligent Transportation Systems, ACM Transactions on Internet Technology, IEEE Transactions on Network Science and Engineering, IEEE Transactions on Multimedia, IEEE Transactions on Computational Social Systems, and IEEE Transactions on Emerging Topics in Computational Intelligence and many top conference papers in the fields of edge intelligence. His main research interests include deep learning for the Internet of Things.
	\end{IEEEbiography}
	\begin{IEEEbiography}[{\includegraphics[width=1in,height=1.25in,clip,keepaspectratio]{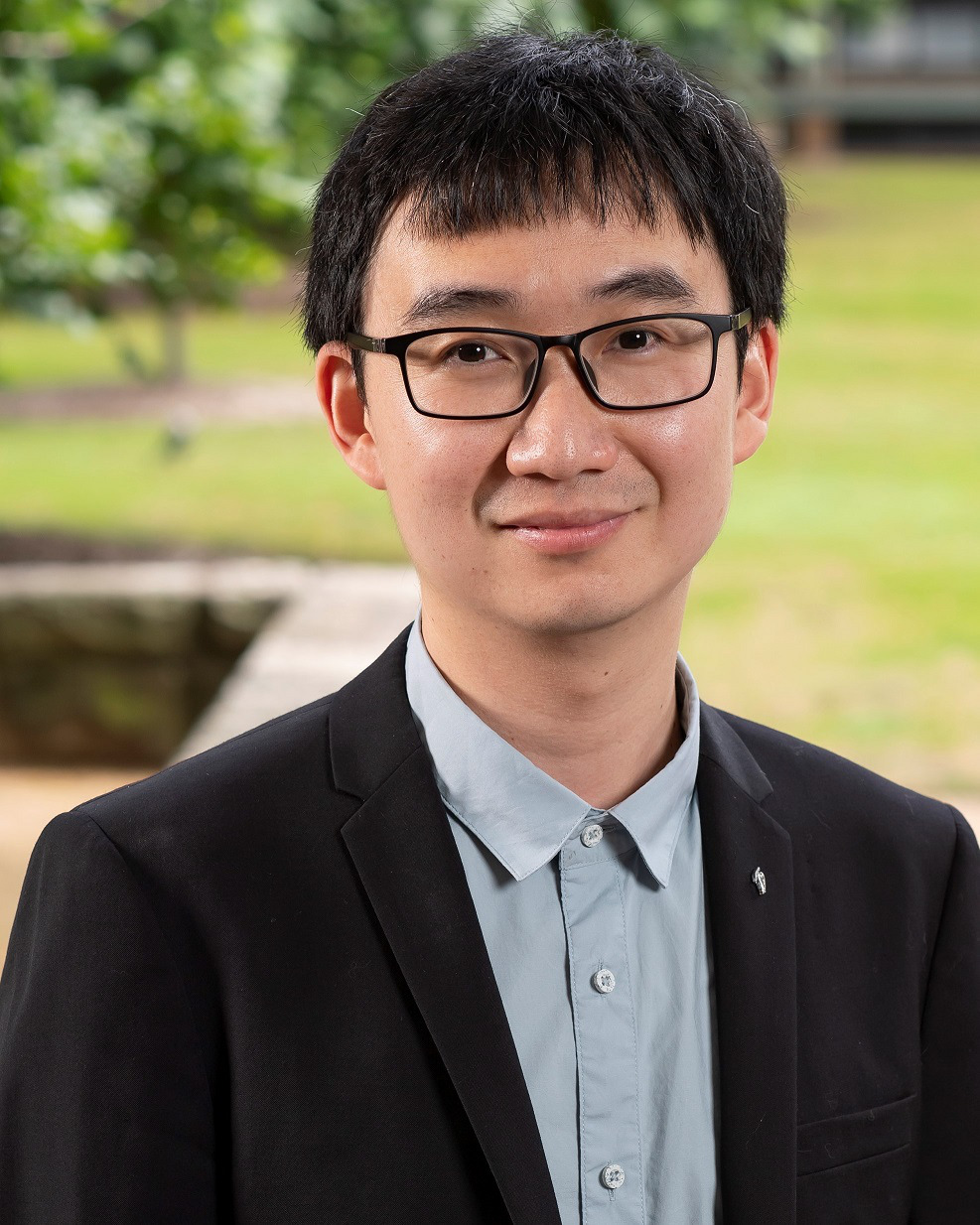}}]
		{Chang Liu} (M'19) received the Ph.D. degree from Dalian University of Technology, China, in 2017. In 2015, he was a joint Ph.D. Scholar with the University of Tennessee, Knoxville, TN, USA. From 2017 to 2022, he was a Post-Doctoral Research Fellow with the University of Electronic Science and Technology of China, and a Research Fellow with the University of New South Wales, Sydney, Australia. Since 2022, he has been an Alexander von Humboldt Research Fellow with the Institute for Digital Communications, Friedrich Alexander University Erlangen-Nuremberg (FAU), Erlangen, Germany. From 2022 to 2023, he worked as a Research Assistant Professor at The Hong Kong Polytechnic University. He is currently a Lecturer (Assistant Professor in U.S. systems) with the Department of Computer Science and Information Technology, La Trobe University, Melbourne, Australia. To date, he has published more than 50 papers in leading international journals and conferences. He was a recipient of the Best Paper Award at the IEEE International Symposium on Wireless Communication Systems in 2022 and an Exemplary Reviewer from IEEE Transactions on Communications. He is a foundation member of IEEE ComSoc special interest group on orthogonal time frequency space (OTFS), a Lead Guest Editor of Future Internet, and a Guest Editor of Frontiers in Communications and Networks. His research interests include machine learning for communications, integrated sensing and communication (ISAC), intelligent reflecting surface (IRS)-assisted communications, OTFS, Internet of Things (IoT), and cognitive radio. 
	\end{IEEEbiography}
	\begin{IEEEbiography}[{\includegraphics[width=1.1in,height=1.25in,clip,keepaspectratio]{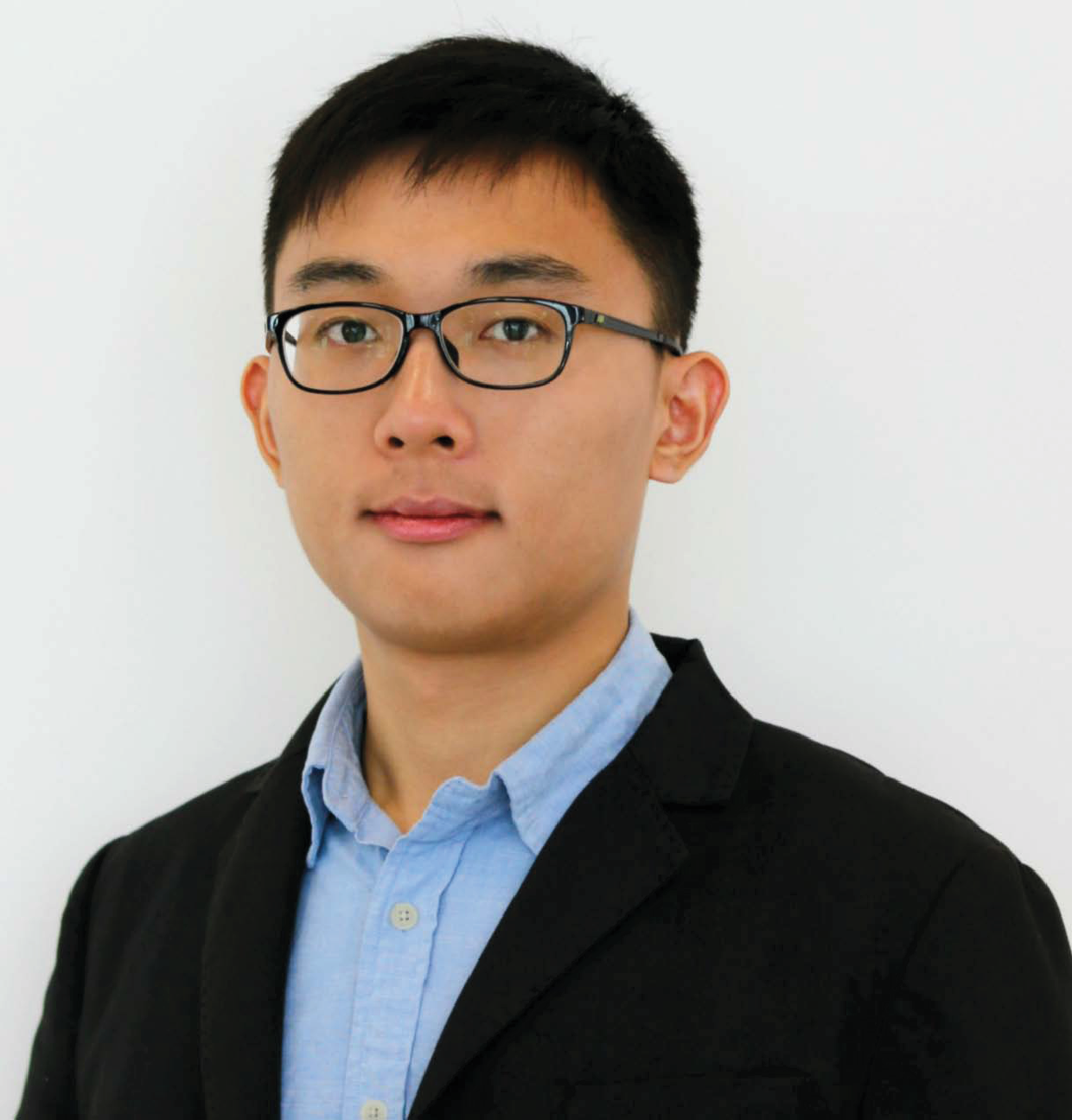}}]
		{Zhiqiang Wei (S'16-M'19)} received the B.E. degree in information engineering from Northwestern Polytechnical University (NPU), Xi'an, China, in 2012, and the Ph.D. degree in electrical engineering and telecommunications from the University of New South Wales (UNSW), Sydney, Australia, in 2019. From 2019 to 2020, he was a Postdoctoral Research Fellow with UNSW. From 2021 to 2022, he was a Humboldt Postdoctoral Research Fellow with the Institute for Digital Communications, Friedrich-Alexander University Erlangen-Nuremberg (FAU), Erlangen, Germany. He is currently a Professor with the School of Mathematics and Statistics, Xi'an Jiaotong University, Xi'an, China. He is the founding co-chair (publications) of the IEEE ComSoc special interest group on OTFS (OTFS-SIG). He received the Best Paper Award at the IEEE ICC 2018 and IEEE WCNC 2023. He was the organizer/chair for several workshops and tutorials on related topics of orthogonal time frequency space (OTFS) in IEEE flagship conferences, including IEEE ICC, IEEE WCNC, IEEE VTC, and IEEE ICCC. He also co-authored the IEEE ComSoc Best Readings on OTFS and Delay Doppler Signal Processing. He is now serving as the Associate Editor of the IEEE Open Journal of the Communications Society. His current research interests include delay-Doppler communications, resource allocation optimization, and statistic and array signal processing.
\end{IEEEbiography}
	\begin{IEEEbiography}[{\includegraphics[width=0.95in,height=1.4in]{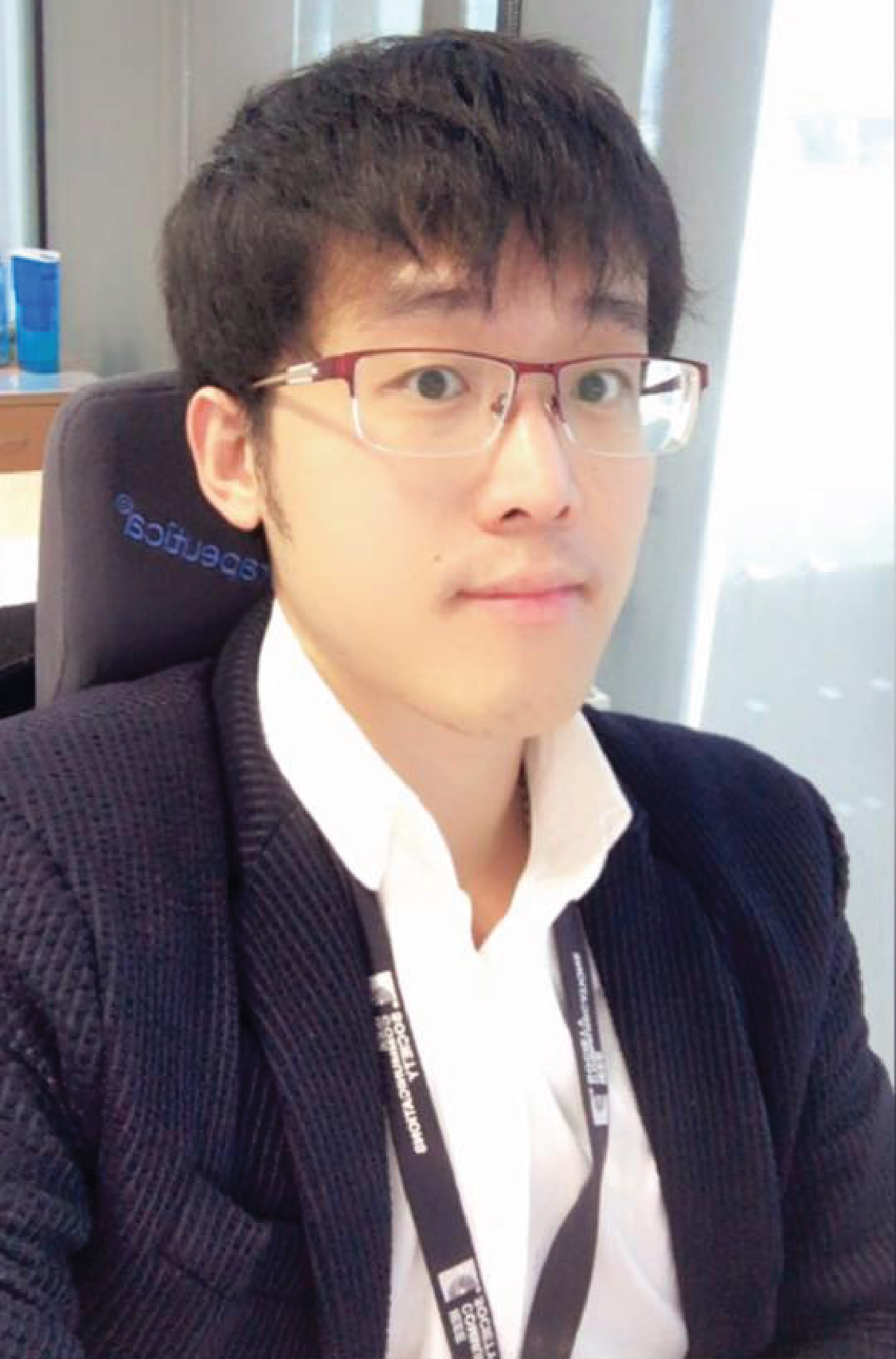}}]
		{Derrick Wing Kwan Ng (S'06-M'12-SM'17-F'21)} received his bachelor's degree (with first-class Honors) and the Master of Philosophy degree in electronic engineering from The Hong Kong University of Science and Technology (HKUST), Hong Kong, in 2006 and 2008, respectively, and his Ph.D. degree from The University of British Columbia, Vancouver, BC, Canada, in November 2012. He was a senior postdoctoral fellow at the Institute for Digital Communications, Friedrich-Alexander-University Erlangen-N\"urnberg (FAU), Germany. He is currently a Scientia Associate Professor with the University of New South Wales, Sydney, NSW, Australia. His research interests include global optimization,  integrated sensing and communication (ISAC),  physical layer security, IRS-assisted communication, UAV-assisted communication, wireless information and power transfer, and green (energy-efficient) wireless communications.

		Since 2018, he has been listed as a Highly Cited Researcher by Clarivate Analytics (Web of Science). He was the recipient of the Australian Research Council (ARC) Discovery Early Career Researcher Award 2017, IEEE Communications Society Leonard G. Abraham Prize 2023, IEEE Communications Society Stephen O. Rice Prize 2022, Best Paper Awards at the WCSP 2020, 2021, IEEE TCGCC Best Journal Paper Award 2018, INISCOM 2018, IEEE International Conference on Communications (ICC) 2018, 2021, 2023, IEEE International Conference on Computing, Networking and Communications (ICNC) 2016, IEEE Wireless Communications and Networking Conference (WCNC) 2012, IEEE Global Telecommunication Conference (Globecom) 2011, 2021, and IEEE Third International Conference on Communications and Networking in China 2008. From January 2012 to December 2019, he served as an Editorial Assistant to the Editor-in-Chief of the IEEE Transactions on Communications. He is also the Editor of the IEEE Transactions on Communications and an Associate Editor-in-Chief for the IEEE Open Journal of the Communications Society.
\end{IEEEbiography}

\end{document}